\documentclass[11pt]{article}
\pdfoutput=1
\usepackage{aas_macros,jcappub,natbib,float,caption,comment,bm}
\usepackage{graphicx,epsfig}
\usepackage[dvipsnames]{xcolor}
\usepackage[utf8]{inputenc}

\def\be{\begin{equation}}
\def\ee{\end{equation}}
\def\ba#1\ea{\begin{align}#1\end{align}}
\newcommand{\vs}{\nonumber\\}

\newcommand{\refeq}[1]{eq.~(\ref{eq:#1})}
\newcommand{\refeqs}[2]{eqs.~(\ref{eq:#1})--(\ref{eq:#2})}
\newcommand{\refEq}[1]{Eq.~(\ref{eq:#1})}
\newcommand{\refEqs}[2]{Eqs.~(\ref{eq:#1})--(\ref{eq:#2})}
\newcommand{\reffig}[1]{figure~\ref{fig:#1}}
\newcommand{\refFig}[1]{Figure~\ref{fig:#1}}

\newcommand{\refsec}[1]{Sec.~\ref{sec:#1}}
\newcommand{\refapp}[1]{App.~\ref{app:#1}}

\renewcommand{\[}{\left[}
\renewcommand{\]}{\right]}
\renewcommand{\(}{\left(}
\renewcommand{\)}{\right)}

\let\oldv\v

\renewcommand{\v}[1]{\mathbf{#1}}

\newcommand{\vq}{\v{q}}
\newcommand{\vk}{\v{k}}

\newcommand{\ihMpc}{h~{\rm Mpc}^{-1}}
\newcommand{\hMpcc}{h^{-3}~{\rm Mpc}^3}

\newcommand{\cO}{\mathcal O}
\newcommand{\lnPp}{\ln P_l'(k)}
\newcommand{\hk}{\hat{k}}
\newcommand{\hq}{\hat{q}}
\newcommand{\hn}{\hat{n}}
\newcommand{\hP}{\hat{P}}
\newcommand{\nbar}{\bar{n}}
\renewcommand{\Re}{{\rm Re}}
\renewcommand{\Im}{{\rm Im}}
\newcommand{\Tau}{\mathcal{T}}

\title{Power spectrum in the presence of large-scale overdensity and tidal fields: breaking azimuthal symmetry}

\author[a]{Chi-Ting Chiang}
\author[b]{and An\oldv{z}e Slosar}

\affiliation[a]{C.N. Yang Institute for Theoretical Physics, Stony Brook University, Stony Brook, NY 11794, U.S.A.}
\affiliation[b]{Physics Department, Brookhaven National Laboratory, Upton NY 11973, USA}

\emailAdd{chi-ting.chiang@stonybrook.edu}
\emailAdd{anze@bnl.gov}

\abstract{We consider the power spectrum of a biased tracer observed
  in a finite volume in the presence of a large-scale overdensity and
  tidal fields. Expanding both the observed power spectrum and the
  source fields (linear power spectrum, scalar overdensity and tidal
  field tensor) in spherical harmonics, we explicitly confirm that
  each $(\ell,m)$ source generates just the corresponding $(\ell,m)$
  modes of power spectrum in real space. In redshift space, each
  $(\ell,m)$ source additionally couples only to $(\ell+2n,m)$ modes
  of tracer power spectra. This generalizes the Kaiser formula for
  monopole, quadrupole and hexadecapole of the power spectrum to all
  $(\ell,m)$ modes generated to the second order in perturbation theory.
  This formalism can find applications in constraining the super-sample
  covariance and in the local power spectrum based bispectrum estimators.
  As an example application, we forecast the ability to measure these
  modes a survey with BOSS-like galaxy number densities.
}

\begin{document}

\subheader{\rm YITP-SB-18-05}

\maketitle
\flushbottom

\section{Introduction}
\label{sec:introduction}

The universe has no global special direction and no global special
place. Statistical isotropy and homogeneity on large scale are two
of the most fundamental assumptions of cosmology. It means that any
statistically non-zero quantity must respect these constraints. The
translational invariance calls for a Fourier-space description of
the two-point correlation, i.e. the power spectrum, and statistical
isotropy additionally requires that the power spectrum is the same
in all directions.

The observed galaxy power spectrum, however, is not isotropic. There
is a special direction, the direction along the line-of-sight, where
galaxies are displaced from their nominal, cosmological redshift-given
coordinate by the component of their peculiar velocities along the
line-of-sight. This results in redshift-space distortions which in
Fourier space introduce additional dependence of the power spectrum
on the cosine of the polar angle with respect to the line-of-sight
$\mu$ \cite{Kaiser:1987qv}. Conventionally, the full anisotropic
power spectrum is expanded in the line-of-sight using the Legendre
polynomials and one finds that the redshift-space distortions, on
top of the monopole, generate the quadrupole and hexadecapole moments
for the Kaiser power spectrum (the odd $\ell$ moments cannot be
generated in auto-correlation due to symmetry along $\mu=0$ line).

However, full azimuthal symmetry with respect to the line-of-sight
still remains, meaning that the power spectrum is independent on the
azimuthal angle $\phi$. In fact, unless there is an additional preferred
direction in the primordial physics (see the discussions of various
models in Ref.~\cite{Shiraishi:2016wec} and the references therein)
which is not aligned with the line-of-sight, the symmetry of the system
does not allow the statistical properties of the observables to depend
on $\phi$. Of course, any real measurement in the finite volume will
produce scatter, which will however be consistent with zero azimuthal
power. 

While it is nontrivial to break the azimuthal symmetry for the entire
universe, for a finite volume the azimuthal symmetry is generally broken,
even without the anisotropic survey window function and in the simplest
cosmological model, i.e. single-field inflation, Einstein gravity, and
$\Lambda$CDM background. Specifically, a finite volume rests in the
background of long-wavelength modes, whose expected variances are given
by the convolution of the power spectrum and the window function (see
\refeqs{sigmad}{sigmaT}). These super-volume scale modes are not directly
observable for the local observer sitting in the volume because the underlying
mean density of the entire universe is unknown, but they do generate both
a mean overdensity and tide, which act as a global constant scalar and
tensor over the volume. Gravitational evolution couples the Fourier modes
of different wavenumbers, so the long-wavelength perturbations will affect
the evolution of the small-scale structure formation (see e.g. Ref.~\cite{Bernardeau:2001qr}
for a review). This immediately calls for a natural extension of the Kaiser
formalism: instead of limiting to the power spectrum $\ell=0,2,4$ multipoles,
we will additionally allow for non-zero $m$ moments of the power spectrum
due to the presence of the super-volume modes. We stress that we are
expanding the tracer power spectrum and \emph{not} the tracer overdensity
field. To maximally exploit the symmetries of the problem, we will do the
same for the source fields, in particular, we will treat tidal tensor as a
general quadrupole \cite{Schmidt:2013gwa,Osato:2017ess} and the overdensity
of the volume and the linear power spectrum as trivial monopoles. This
will simplify the treatment considerably.

The effect of the long-wavelength modes on the power spectrum has been
studied extensively for the overdensity
\cite{Li:2014sga,Li:2014jra,Chiang:2014oga,Wagner:2015gva,Chiang:2017jnm,Dai:2015jaa,Barreira:2017sqa}
as well as the tidal fields
\cite{Schmidt:2013gwa,Nishimichi:2015kla,Akitsu:2016leq,Akitsu:2017syq,Li:2017qgh,Ip:2016jji,Schmidt:2018hbj,Pen:2012ft,Zhu:2015zlh,Zhu:2016esh}.
Many of the studies focus on the effect of the covariance due to the
super-sample mode, or the super-sample covariance, and the constraints
on parameters \cite{Li:2014sga,Akitsu:2016leq,Akitsu:2017syq,Li:2017qgh}.
Ref.~\cite{Li:2014jra} has treated the super-sample overdensity as a
signal and discuss the possible constraint. Ref.~\cite{Li:2017qgh}
studied this effect as contamination to baryonic acoustic oscillations
(BAO) and redshift-space distortions measurements and included all bias
parameters. They used the full bias parameterization to the second order
and test it on simulations. However, because they focused on the effect
as contamination, only the usual, azimuthally averaged power spectrum
was considered. Ref.~\cite{Akitsu:2017syq} has also treated the large-scale
tidal field both as a contamination and as a signal, but only for azimuthally
averaged component and did not include the higher-order biases. Ref.~\cite{Shiraishi:2016wec}
did go beyond azimuthally averaged power spectrum, but treated the power
asymmetry as a statistical field using the spherical bipolar formalism.
In Ref.~\cite{Schmidt:2018hbj} the effect has been studied for dark matter
in simulations using the response function. This work has confirmed that
the tidal effect, at least for dark matter, never significantly exceeds
the large-scale perturbation theory. We will later use the same intuition
to argue that non-azimuthally symmetric components of the anisotropic
power spectrum are unlikely to be contaminated by nonlinearities.

The goal of this paper is to generalize the discussion and explore
observability of all components of the tidal fields from the
redshift-space galaxy power spectrum of a finite survey. To that
extend, we will follow the standard derivation of the power spectrum
response along the lines of Ref.~\cite{Akitsu:2017syq}, but including
all the bias terms as in Ref.~\cite{Li:2017qgh}. However, rather than
azimuthally averaging the resulting expression we will expand it in
spherical harmonic base.

The rest of the paper is organized as follows.
In \refsec{theory} we derive the power spectrum in the presence of long-wavelength
overdensity and tidal fields, and decompose the power spectrum using the spherical
harmonics to highlight the internal symmetry of the problem. 
In \refsec{fisher} we use Fisher analysis to forecast the constraints on the
long modes for a finite survey, and study the degeneracies between the linear
bias as well as the growth rate with the long modes.
We conclude in \refsec{conclusion}.
In \refapp{Amn} we show explicitly the terms with different angular dependencies
of the redshift-space power spectrum in the presence of long-wavelength overdensity
and tidal fields.

\section{Theory}
\label{sec:theory}

\subsection{Power spectrum in a volume with of a long-wavelength density perturbation}
\label{sec:longmode}
Consider measuring galaxy power spectrum in a volume $V$, which is large enough
to encompass linear modes over some range of scales. Within this volume the mean matter
density perturbation $\delta^W$ and tide $\tau^W_{ij}$ are given by
\be
 \delta^W=\int\frac{d^3k}{(2\pi)^3} W(-\vk)\delta(\vk) \,, \quad
 \tau^W_{ij}=\int\frac{d^3k}{(2\pi)^3} W(-\vk)\(\hk_i\hk_j-\frac{1}{3}\delta^K_{ij}\)\delta(\vk) \,,
\label{eq:longmodes}
\ee
where $W(\vk)$ is the top-hat smoothing kernel, $\delta^K_{ij}$ is the Kronecker
delta, and hat refers to the unit vector. For simplicity we shall assume the
smoothing kernel is isotropic and so $W(\vk)=W(k)$.
Even though for a single mode $\tau_{ij}(\vk)$ is deterministically determined
from $\delta(\vk)$, after averaging over all the modes inside the volume $V$,
knowing $\delta^W$, which is a scalar containing just a number, is not sufficient
to predict $\tau^W_{ij}$, which is a symmetric and traceless tensor containing
five numbers, since they have different $k$ weightings according to \refeq{longmodes}.
This is particularly crucial when considering the super-sample modes as the underlying
$\delta(\vk)$ is unknown given that we have only a finite survey. Thus, in this
paper we shall consider $\delta^W$ and $\tau^W_{ij}$ as separate variables, meaning
that when studying the constraints on the long modes we have to constrain $\delta^W$
and $\tau^W_{ij}$ separately.
Due to the presence of the long modes, the power spectrum in this volume $V$
will be affected as
\cite{Li:2014sga,Li:2014jra,Chiang:2014oga,
Wagner:2015gva,Chiang:2017jnm,Nishimichi:2015kla,Akitsu:2016leq,Akitsu:2017syq,
Li:2017qgh,Dai:2015jaa,Ip:2016jji,Barreira:2017sqa,Schmidt:2018hbj}
\be
 P_{gg}(\vk|\delta^W,\tau^W_{ij})=P_{gg}(\vk)+\frac{dP_{gg}(\vk)}{d\delta^W}\delta^W
 +\sum_{ij}\frac{dP_{gg}(\vk)}{d\tau^W_{ij}}\tau^W_{ij}
 +\cO\[\(\delta^W\)^2,\(\tau^W_{ij}\)^2,\delta^W\tau^W_{ij}\] \,,
\ee
hence the power spectrum contains additional six degrees of freedom: one from
$\delta^W$ and five from $\tau^W_{ij}$ due to the symmetric and traceless conditions.
On average $\langle P_{gg}(\vk|\delta^W,\tau^W_{ij})\rangle=P_{gg}(\vk)$
since $\langle\delta^W\rangle=\langle\tau^W_{ij}\rangle=0$, but if we correlate
the power spectrum with the long modes in the same volume as for measuring
the position-dependent power spectrum \cite{Chiang:2014oga,Chiang:2015eza},
then one would pick up the response signal.

The response of the power spectrum to the long mode is equivalent to the bispectrum
in the squeezed limit \cite{Chiang:2014oga,Wagner:2015gva,Barreira:2017sqa}.
Intuitively, the squeezed bispectrum measures the correlation between one long
and two short modes, and one can regard the two short modes as the small-scale
power spectrum and the long mode as the large-scale perturbation. Since we consider
the response of redshift-space galaxy power spectrum to the underlying long node,
we shall adopt the bispectrum formed by two small-scale redshift-space galaxy
perturbations and one large-scale real-space matter perturbation. Specifically,
\ba
 \langle\delta^W P_{gg}(\vk|\delta^W,\tau^W_{ij})\rangle
 \:&=\frac{dP_{gg}(\vk)}{d\delta^W}\langle(\delta^W)^2\rangle
 +\sum_{ij}\frac{dP_{gg}(\vk)}{d\tau^W_{ij}}\langle\delta^W\tau^W_{ij}\rangle \vs
 \:&=\lim_{q/k\ll1}B_{mgg}(\vq,\vk,-\vq-\vk)
 =B^{\rm sq}_{mgg}(\vq,\vk,-\vq-\vk) \,,
\label{eq:Bmgg_0}
\ea
where for simplicity we consider $\delta^W$ and $\tau^W_{ij}$ to contain a single
mode with wavelength $\vq$. Therefore, by the squeezed bispectrum prescription,
we can read off the galaxy power spectrum response to $\delta^W$ and $\tau^W_{ij}$.

The galaxy redshift-space bispectrum predicted by the standard perturbation theory
at the tree-level is \cite{Bernardeau:2001qr}
\be
 B_{ggg}(\vk_1,\vk_2,\vk_3)=2\[Z_1(\vk_1)Z_1(\vk_2)Z_2(\vk_1,\vk_2)P_l(k_1)P_l(k_2)+(2~{\rm cyclic})\] \,,
\ee
where $P_l$ is the linear power spectrum and $Z_1$ and $Z_2$ are the
redshift-space kernels given by
\ba
 Z_1(\vk_i)=\:&b_1+f\mu_{k_i}^2 \,, \vs
 Z_2(\vk_1,\vk_2)=\:&b_1F_2(\vk_1,\vk_2)+\frac{b_2}{2}+\frac{b_{s^2}}{2}S_2(\vk_1,\vk_2)
 +f\mu_{k_3}^2G_2(\vk_1,\vk_2) \vs
 \:&-\frac{f\mu_{k_3}k_3}{2}\[\frac{\mu_{k_1}}{k_1}(b_1+f\mu_{k_1}^2)+\frac{\mu_{k_2}}{k_2}(b_1+f\mu_{k_2}^2)\] \,.
\ea
Here, $b_1$, $b_2$, and $b_{s^2}$ are linear, nonlinear, and tidal biases,
$f$ is the growth rate, $\mu_{k_i}$ is the cosine between $\vk_i$ and the
line-of-sight, and
\ba
 F_2(\vk_1,\vk_2)\:&=\frac{5}{7}+\frac{\mu_{k_1k_2}}{2}\(\frac{k_1}{k_2}+\frac{k_2}{k_1}\)
 +\frac{2}{7}\mu_{k_1k_2}^2 \vs
 G_2(\vk_1,\vk_2)\:&=\frac{3}{7}+\frac{\mu_{k_1k_2}}{2}\(\frac{k_1}{k_2}+\frac{k_2}{k_1}\)
 +\frac{4}{7}\mu_{k_1k_2}^2 \vs
 S_2(\vk_1,\vk_2)\:&=\mu_{k_1k_2}^2-\frac{1}{3} \,,
\ea
with $\mu_{k_1k_2}$ being the cosine between $\vk_1$ and $\vk_2$. Using the
redshift-space kernel, the squeezed bispectrum formed by two small-scale
redshift-space galaxy perturbations and one large-scale real-space matter
perturbation is
\ba
 \:&B^{\rm sq}_{mgg}(\vq,\vk,-\vq-\vk) \vs
 =\:&2\[Z_1(\vk)Z_2(\vq,\vk)P_l(q)P_l(k)+Z_1(-\vq-\vk)Z_2(\vq,-\vq-\vk)P_l(q)P_l(|\vq+\vk|)\] \vs
 =\:&\Bigg[\frac{13}{7}b_1^2+2b_1b_2-\frac{2}{3}b_1b_{s^2}+\frac{18}{7}b_1f\mu_k^2+2b_1^2f\mu_k^2 \vs
 \:&~~~+2b_2f\mu_k^2-\frac{2}{3}b_{s^2}f\mu_k^2+\frac{5}{7}f^2\mu_k^4+2b_1f^2\mu_k^4\Bigg]P_l(q)P_l(k) \vs
 +\:&\Bigg[\frac{8}{7}b_1^2+2b_1b_{s^2}-b_1^2\lnPp+\frac{24}{7}b_1f\mu_k^2+2b_{s^2}f\mu_k^2 \vs
 \:&~~~-2b_1f\mu_k^2\lnPp+\frac{16}{7}f^2\mu_k^4-f^2\mu_k^4\lnPp\Bigg]\mu_{kq}^2P_l(q)P_l(k) \vs
 +\:&\Bigg[-b_1^2f\mu_k\lnPp+4b_1f^2\mu_k^3-2b_1f^2\mu_k^3\lnPp \vs
 \:&~~~+4f^3\mu_k^5-f^3\mu_k^5\lnPp\Bigg]\mu_{kq}\mu_qP_l(q)P_l(k) \vs
 +\:&(b_1^2f-f^3\mu_k^4)\mu_q^2P_l(q)P_l(k)+\cO\(q/k\) \,,
\label{eq:Bmgg}
\ea
where prime is the logarithmic derivative with respect to $k$, $\vq$ and $\vk$
are the long and short modes, and we take $q/k\ll1$. Note that \refeq{Bmgg}
has been derived in Ref.~\cite{Li:2017qgh}, with a slightly different notation.

For point tracers, there will be additional term associated with the
Poissonian process. Namely, in the presence of the large-scale mode
$\vq$, the local number density will be modulated as $\nbar(1+b_1\delta^W)$
with $\nbar$ being the global mean number density, leading to local
modulation of shot-noise term, which will add a $-b_1\nbar^{-1} P_l(q)$
to \refeq{Bmgg}. This is a real term in the bispectrum, but since we
are eventually interested in the locally measured power spectrum for
which the shot-noise term will be locally subtracted, we will dismiss
it in this paper.
In addition, the presence of the large-scale tide would also cause
modulation in local galaxy number density, though the leading-order
effect is second order.
The impact of long modes on stochasticity is discussed in detailed
in Sec.~2.8 of \cite{Desjacques:2016bnm}.

To extract the power spectrum response from \refeq{Bmgg}, we first note
that the large-scale tide is related to the large-scale overdensity as
\be
 \tau^W_{ij}=\(\hq_i\hq_j-\frac{1}{3}\delta^K_{ij}\)\delta^W \,.
\label{eq:tauij}
\ee
This allows us to write
\ba
 \:&\mu_{kq}^2=\sum_{ij}\hk_i\hk_j\hq_i\hq_j=\frac{1}{3}+\frac{1}{\delta^W}\sum_{ij}\hk_i\hk_j\tau^W_{ij} \,, \vs
 \:&\mu_{kq}\mu_q=\sum_{ij}\hk_i\hn_j\hq_i\hq_j=\frac{\mu_k}{3}+\frac{1}{\delta^W}\sum_{ij}h_{ij}\tau^W_{ij} \,, \vs
 \:&\mu_q^2=\sum_{ij}\hn_i\hn_j\hq_i\hq_j=\frac{1}{3}+\frac{1}{\delta^W}\sum_{ij}\hn_i\hn_j\tau^W_{ij} \,,
\label{eq:angles_tauij}
\ea
where $\hn$ is the line-of-sight unit vector and $h_{ij}=(\hk_i\hn_j+\hk_j\hn_i)/2$.
Plugging \refeq{angles_tauij} into \refeq{Bmgg} and using the fact that
the power spectrum of the long mode can be regarded as $P_l(q)\sim(\delta^W)^2$,
the power spectrum responses can be read off by comparing terms with \refeq{Bmgg_0}.
Specifically, we have the galaxy power spectrum response to $\delta^W$ as
\ba
 \:&\frac{1}{P_l(k)}\frac{dP_{gg}(\vk)}{d\delta^W} \vs
 =\:&\frac{47}{21}b_1^2+2b_1b_2-\frac{1}{3}b_1^2\lnPp+\frac{1}{3}b_1^2f+\frac{26}{7}b_1f\mu_k^2
 +2b_1^2f\mu_k^2+2b_2f\mu_k^2 \vs
 -\:&\frac{2}{3}b_1f\mu_k^2\lnPp-\frac{1}{3}b_1^2f\mu_k^2\lnPp+\frac{31}{21}f^2\mu_k^4
 +\frac{10}{3}b_1f^2\mu_k^4-\frac{1}{3}f^2\mu_k^4\lnPp \vs
 -\:&\frac{2}{3}b_1f^2\mu_k^4\lnPp-\frac{1}{3}f^3\mu_k^4+\frac{4}{3}f^3\mu_k^6-\frac{1}{3}f^3\mu_k^6\lnPp \,,
\label{eq:delta_resp}
\ea
and to $\tau_{ij}^W$ as
\ba
 \:&\frac{1}{P_l(k)}\frac{dP_{gg}(\vk)}{d\tau^W_{ij}} \vs
 =\:&\Bigg[\frac{8}{7}b_1^2+2b_1b_{s^2}-b_1^2\lnPp+\frac{24}{7}b_1f\mu_k^2+2b_{s^2}f\mu_k^2
 -2b_1f\mu_k^2\lnPp+\frac{16}{7}f^2\mu_k^4-f^2\mu_k^4\lnPp\Bigg]\hk_i\hk_j \vs
 +\:&\[-b_1^2f\mu_k\lnPp+4b_1f^2\mu_k^3-2b_1f^2\mu_k^3\lnPp+4f^3\mu_k^5-f^3\mu_k^5\lnPp\]h_{ij} \vs
 +\:&\[b_1^2f-f^3\mu_k^4\]\hn_i\hn_j \,.
\label{eq:tauij_resp}
\ea
\refEq{delta_resp} and \refeq{tauij_resp} are basically the same as in
Ref.~\cite{Nishimichi:2015kla} and Ref.~\cite{Akitsu:2017syq} respectively,
except the addition bias parameters. It is useful to decompose the redshift-space
galaxy power spectrum and response into different angular dependencies as
\ba
 \:&P_{gg}(\vk)=\sum_{n=0}^2A_{0,n}\mu_k^{2n} \,, \quad
 \frac{dP_{gg}(\vk)}{d\delta^W}=\sum_{n=0}^3A_{1,n}(k)\mu_k^{2n} \,, \vs
 \:&\frac{dP_{gg}(\vk)}{d\tau^W_{ij}}=\sum_{n=0}^2A_{2,n}(k)\mu_k^{2n}\hk_i\hk_j
 +\sum_{n=0}^2A_{3,n}(k)\mu_k^{2n+1}h_{ij}+\sum_{n=0}^1A_{4,n}(k)\mu_k^{4n}\hn_i\hn_j \,,
\ea
where $A_{m,n}$ are given explicitly in \refapp{Amn}.

The above calculation assumes that the underlying mean galaxy number density
is known for the power spectrum calculation. This is true for the subvolumes
inside a survey because the underlying mean number density can be computed
from the survey, assuming that the super-volume modes (larger than the survey)
have negligible impact. On the other hand, to extract the super-volume modes,
the underlying mean number density requires the observation of even larger
volume (in principle the whole universe) so is generally unknown, and one can
only use the ``local'' mean density in the survey to characterize the power
spectrum. We refer to this as the ``local'' power spectrum, which is related
to the ``global'' power spectrum, computed using the true underlying mean
density, as
\be
 P_{gg}^G(\vk)=(1+\delta_g^W)^2P_{gg}^L(\vk) \,,
\ee
where the superscripts $G$ and $L$ denote respectively the global and
local power spectrum, and $\delta_g^W$ is the mean galaxy overdensity
of the volume. In real space $\delta_g^W=b_1\delta^W$, and in redshift
space
\be
 \delta_g^W=(b_1+f\mu_q^2)\delta^W
 =\(b_1+\frac{1}{3}f\)\delta^W+f\sum_{ij}\hn_i\hn_j\tau^W_{ij}
 =\(b_1+\frac{1}{3}f\)\delta^W+f\tau^W_{22} \,,
\ee
where we conventionally set $\hn=\hat{z}$. Thus, the local and global
power spectra to the leading order are related in real and redshift
space respectively as
\be
 P_{gg}^L(\vk)\approx(1-2b_1\delta^W)P_{gg}^G(\vk) \,, \quad
 P_{gg}^L(\vk)\approx\[1-2\(b_1+\frac{1}{3}f\)\delta^W-2f\tau^W_{22}\]P_{gg}^G(\vk) \,.
\label{eq:local}
\ee
We can use \refeq{local} to mimic the power spectrum measured by a local
observer living in the volume who cannot access $\delta^W$ and $\tau^W_{ij}$.
The same effect has been discussed in Ref.~\cite{Chiang:2017qoh} for probing
the correlation of Cosmic Microwave Background (CMB) lensing convergence
and Lyman-$\alpha$ forest power spectrum measured using the local mean flux.

\subsection{Spherical expansion}
\label{sec:expansion}
To highlight the internal symmetry of the problem, we expand the large-scale
tidal field in the $\ell=2$ spherical harmonics as
\be
 \Tau^W_m=\int d^2\hk\hk_i\hk_j\tau^W_{ij}Y_{2m}^*(\hk) \,.
\ee
The existing components are
\ba
 \:&\Tau^W_0=-\sqrt{\frac{4\pi}{5}}(\tau^W_{00}+\tau^W_{11}) \,, \quad
 \Tau^W_1=-\sqrt{\frac{8\pi}{15}}(\tau^W_{02}-i\tau^W_{12}) \,, \quad
 \Tau^W_2=\sqrt{\frac{2\pi}{15}}(\tau^W_{00}-\tau^W_{11}-2i\tau^W_{01}) \,,
\ea
hence the tidal tensor can be written as
\be
 \tau^W = \sqrt{\frac{15}{2\pi}}\left[ {\begin{array}{ccc}
   -\frac{1}{2\sqrt{6}}\Tau^W_0 + \frac{1}{2} \Re \Tau^W_2 & -\frac{1}{2} \Im \Tau^W_2 & -\frac{1}{2}\Re \Tau^W_1 \\
   -\frac{1}{2} \Im \Tau^W_2 & -\frac{1}{2\sqrt{6}}\Tau^W_0 - \frac{1}{2} \Re \Tau^W_2 & \frac{1}{2} \Im \Tau^W_1 \\
   -\frac{1}{2}\Re \Tau^W_1  & \frac{1}{2}\Im \Tau^W_1 & \frac{1}{\sqrt{6}} \Tau^W_0 \\
  \end{array} } \right] \,.
\label{eq:tauijmat}
\ee
Note that $\Tau^W_m$ has five degrees of freedom (one real $m=0$ and two
$m=1,2$ complex numbers), matching the number of degrees of freedom in
$\tau^W_{ij}$ as $\tau^W_{ij}$ is symmetric and traceless. We also have
the usual reality condition $\Tau^W_m=(\Tau^W_{-m})^*$. 

In the same spirit we expand the redshift-space galaxy power spectrum
and the response in spherical harmonics. Redshift space-distortions
introduce a special direction. For a large survey the line-of-sight
direction varies at different angular positions. For simplicity in
this paper we will apply the plane-parallel approximation and conventionally
set $\hn=\hat{z}$. The spherical multipole expansion of the power
spectrum is therefore given by
\be
 P_{gg,\ell m} (k) = \int d^2\hk P_{gg}(\vk) Y^*_{\ell m} (\hk) \,.
\ee
The similar decomposition has been proposed by Refs.~\cite{Shiraishi:2016wec,Sugiyama:2017ggb}.
Note that $P_{gg,00}(k)$, $P_{gg,20}(k)$ and $P_{gg,40}(k)$ are the
usual azimuthally averaged redshift-space monopole, quadrupole and
hexadecapole of the power spectra, up to different prefactors.
Note also that for $m\neq0$, $P_{gg,\ell m}$ would contain complex
components as for $\Tau^W_m$.

For $m=0$, linear Kaiser power spectrum (only to $\ell\le4$), $\delta^W$
and $\Tau^W_0$ contribute:
\ba
 P_{gg,00}(k)=\:&\frac{2\sqrt{\pi}}{105}\Bigg[(105A_{0,0}+35A_{0,1}+21A_{0,2})
 +\delta^W(105A_{1,0}+35A_{1,1}+21A_{1,2}+15A_{1,3}) \vs
 \:&\hspace{1cm}+\sqrt{\frac{5}{4\pi}}\Tau^W_0(14A_{2,1}+12A_{2,2}+105A_{4,0}+21A_{4,1})\Bigg] \,, \vs
 P_{gg,20}(k)=\:&\frac{2}{21}\sqrt{\frac{\pi}{5}}\Bigg[(14A_{0,1}+12A_{0,2})+
 \delta^W(14A_{1,1}+12A_{1,2}+10A_{1,3}) \vs
 \:&\hspace{1.2cm}+\sqrt{\frac{5}{4\pi}}\Tau^W_0(21A_{2,0}+11A_{2,1}+9A_{2,2}+12A_{4,1})\Bigg] \,, \vs
 P_{gg,40}(k)=\:&\frac{8\sqrt{\pi}}{1155}
 \[22A_{0,2}+\delta^W(22A_{1,2}+30A_{1,3})+\sqrt{\frac{5}{4\pi}}\Tau^W_0(33A_{2,1}+34A_{2,2}+22A_{4,1})\] \,, \vs
 P_{gg,60}(k)=\:&\frac{16}{231}\sqrt{\frac{\pi}{13}}
 \[\delta^W(2A_{1,3})+\sqrt{\frac{5}{4\pi}}\Tau^W_0(3A_{2,2})\] \,.
\label{eq:Pggm0}
\ea
For $m=1$ we find:
\ba
 P_{gg,21}(k)=\:&\frac{1}{42}
 \Tau^W_1(42A_{2,0}+18A_{2,1}+10A_{2,2}+21A_{3,0}+9A_{3,1}+5A_{3,2}) \,, \vs
 P_{gg,41}(k)=\:&\frac{\sqrt{6}}{231}
 \Tau^W_1(22A_{2,1}+20A_{2,2}+11A_{3,1}+10A_{3,2}) \,, \vs
 P_{gg,61}(k)=\:&\frac{4}{33}\sqrt{\frac{5}{91}}
 \Tau^W_1(2A_{2,2}+A_{3,2}) \,.
\label{eq:Pggm1}
\ea
For $m=2$ we find:
\ba
 P_{gg,22}(k)=\:&\frac{1}{21}\Tau^W_2(21A_{2,0}+3A_{2,1}+A_{2,2}) \,, \vs
 P_{gg,42}(k)=\:&\frac{2\sqrt{3}}{231}\Tau^W_2(11A_{2,1}+6A_{2,2}) \,, \vs
 P_{gg,62}(k)=\:&\frac{16}{33}\sqrt{\frac{1}{182}}\Tau^W_2A_{2,2} \,.
\label{eq:Pggm2}
\ea \refEqs{Pggm0}{Pggm2} are the main results of this paper. We find
that with $\Tau^W_1$ and $\Tau^W_2$, the power spectrum has components
with $m=1,2$, meaning that the azimuthal symmetry of the power
spectrum is broken. This opens a new window for measuring the
super-volume tide from the small-scale power spectrum in a volume:
while the linear Kaiser power spectrum dominates the $m=0$ components
and so $\delta^W$ and $\Tau^W_0$ may be difficult to extract, $m=1,2$
components can only be generated by the tidal fields hence any
measurement is pure signal. In reality, however, the anisotropic
window function will also contaminate the signal in $m=1,2$
\cite{Sugiyama:2017ggb}, and so has to be carefully accounted for. In
principle, gravitational lensing is also likely to generate $m\neq0$
modes, but these will be small for survey-size volumes.

\begin{figure}[t]
\centering
\includegraphics[width=0.495\textwidth]{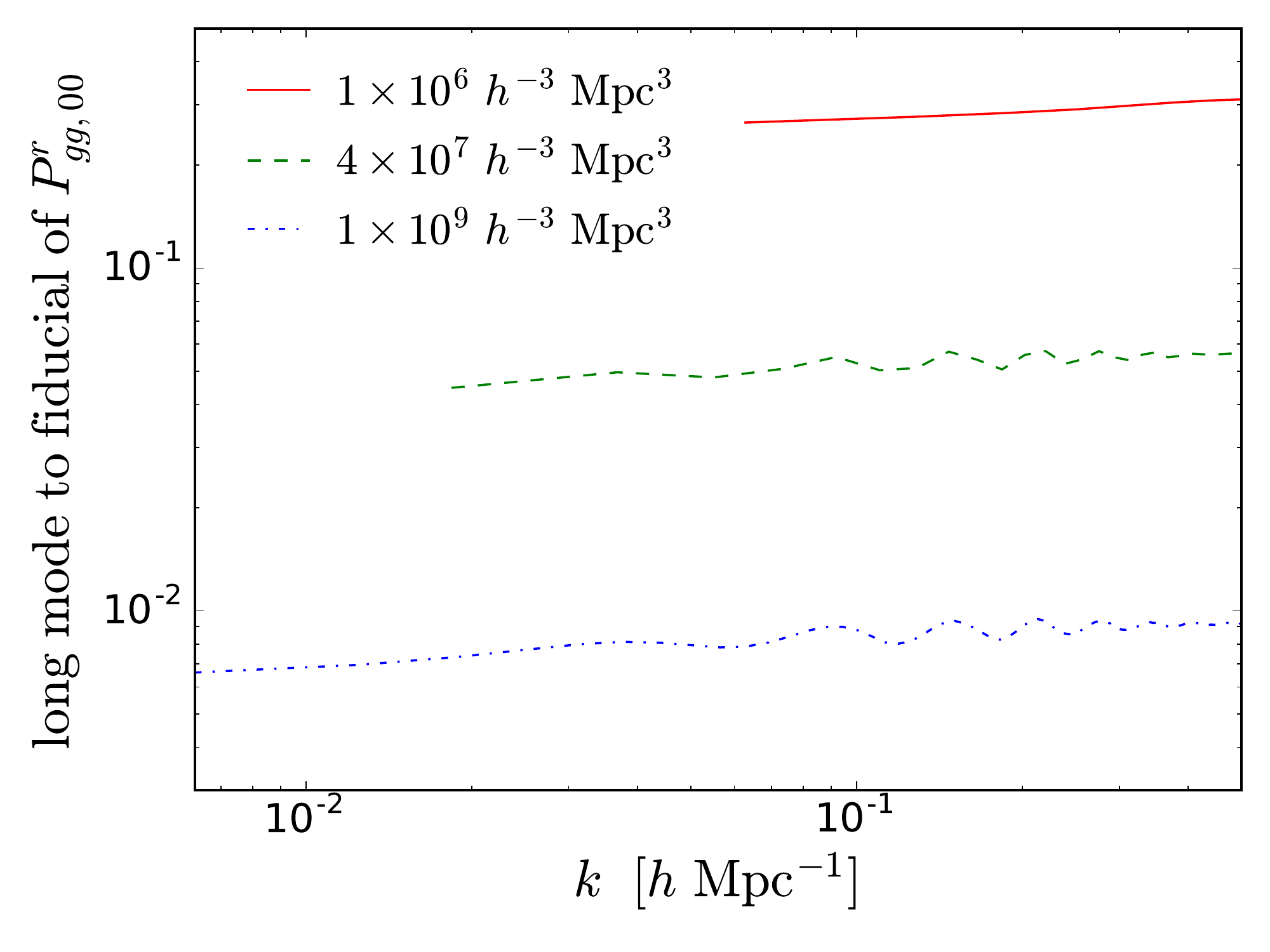}
\includegraphics[width=0.495\textwidth]{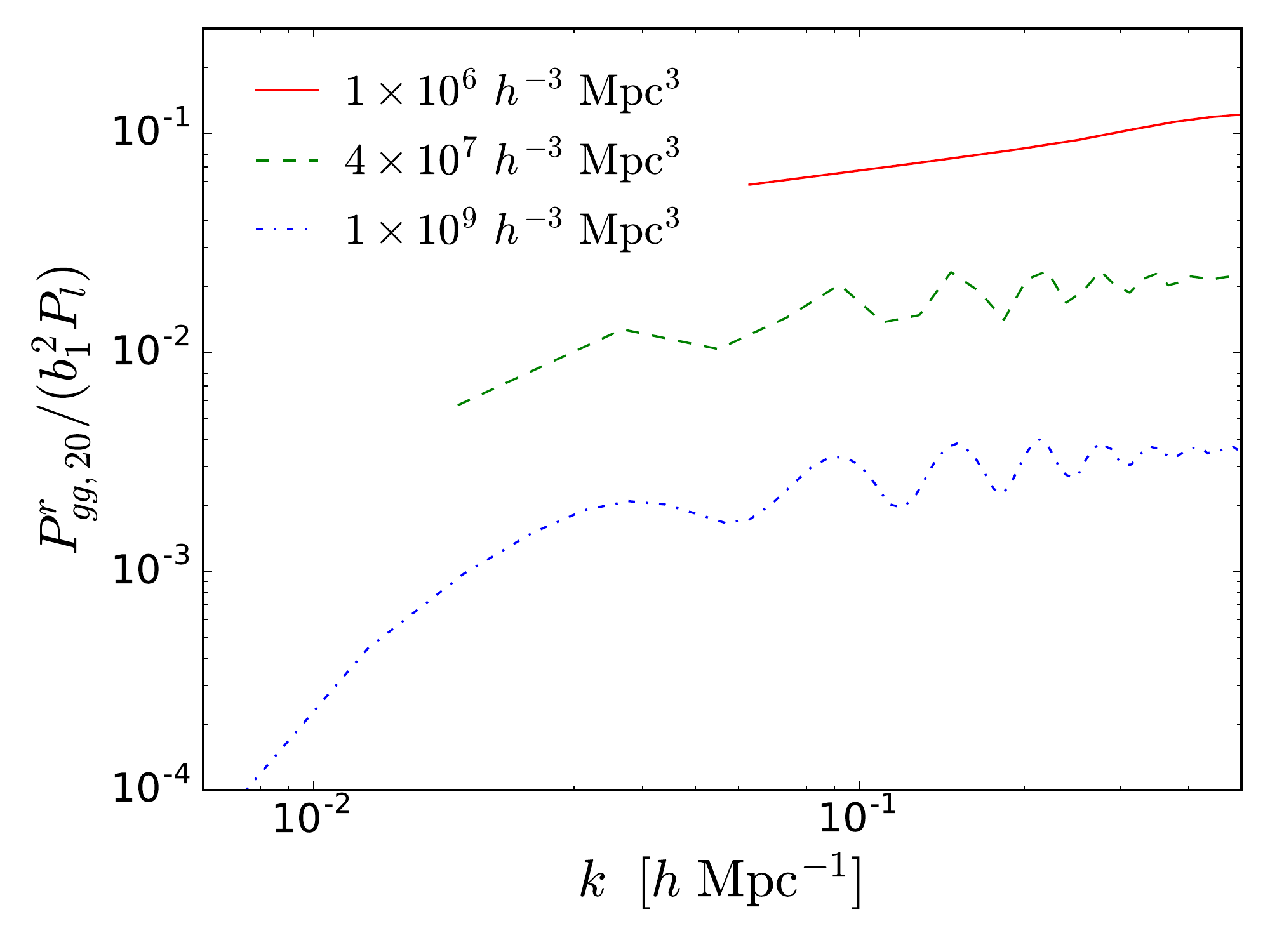}
\caption{(Left) Ratio of the long mode contribution (terms associated with
$\delta^W$) to fiducial ($\delta^W=0$) $P^r_{gg,00}$, which is $b_1^2P_l$
and independent of volume, in real space at $z=0.5$. (Right) Ratio of
$P^r_{gg,20}$ to $b_1^2P_l$ in real space at $z=0.5$. The red solid, green
dashed, and blue dot-dashed lines show volumes of $10^6$, $4\times10^7$,
and $10^9~\hMpcc$. The values of $\delta^W$ and $\Tau^W_0$ are set to be
their $1-\sigma$ expected values, i.e. $\sigma_{\delta^W}$ and $\sigma_{\Tau^W_0}$
with spherical top-hat window function, respectively. Note that the effect
of using the local mean density to compute the power spectrum, i.e. \refeq{local}
is not included.}
\label{fig:Pggr}
\end{figure}

To obtain real-space results, we can set $f=0$ in $A_{m,n}$, and the only
existing components are
\ba
 P^r_{gg,00}(k)=\:&2\sqrt{\pi}\Bigg\lbrace\[b_1^2+\delta^W\(\frac{47}{21}b_1^2+2b_1b_2\)\]P_l(k)
 +\delta^W\(-\frac{1}{3}b_1^2\)P'_l(k)\Bigg\rbrace \,, \vs
 P^r_{gg,2m}(k)=\:&\Tau^W_m\[\(\frac{8}{7}b_1^2+2b_1b_{s^2}\)P_l(k)+\(-b_1^2\)P'_l(k)\] \,. \vs
\label{eq:Pggr}
\ea
These equations have the behavior expected based on the symmetry properties
of the sources: scalar sources give raise to $\ell=0$ modes and tidal $\ell=2,m$
sources give raise to $\ell=2,m$ components of the power spectrum. 

The left panel of \reffig{Pggr} shows the long mode contribution (terms associated
with $\delta^W$) to fiducial ($\delta^W=0$ so independent of volume) $P^r_{gg,00}$,
which is $b_1^2P_l$, in real space at $z=0.5$ for various volumes denoted by
different colors and styles. Note that the minimum wavenumber and the density of
the line reflect the corresponding volume. We set $\delta^W$ to be the $1-\sigma$
expected value, i.e.
\be
 \sigma_{\delta^W}=\langle(\delta^W)^2\rangle^{1/2}=\[\int d^3k |W(k)|^2P_l(k)\]^{1/2} \,,
\label{eq:sigmad}
\ee
where we choose the window function to be spherical top-hat. We find that the
long mode contribution is larger for smaller volume, which is the outcome of
larger $\sigma_{\delta^W}$. We also find that the ratio is fairly scale independent,
hence $\delta^W$ will degenerate with $b_1$ when performing parameter constraint
and we will discuss this in details in \refsec{fisher}. The right panel of \reffig{Pggr}
shows the ratio of $P^r_{gg,20}$ to $b_1^2P_l$ in real space at $z=0.5$, and
as for $\delta^W$ we set the value of $\Tau^W_0$ to be its $1-\sigma$ expected
value, which is
\be
 \sigma_{\Tau^W_0}=\sqrt{2}\sigma_{\Tau^W_1}=\sqrt{2}\sigma_{\Tau^W_2}=\frac{4\sqrt{\pi}}{15}\sigma_{\delta^W} \,.
\label{eq:sigmaT}
\ee
Note that we only show the result for $P^r_{gg,20}$ because it has the same scale
dependence as $P^r_{gg,21}$ and $P^r_{gg,22}$. Compared to the long mode contribution
in $P^r_{gg,00}$, the signal for $P^r_{gg,20}$ is smaller, ranging from $\sim10^{-4}$
to $\sim10^{-1}$ for $V=10^9$ to $10^6~\hMpcc$. However, since the fiducial power
spectrum does not contribute in $P^r_{gg,20}$, any detection of $P^r_{gg,20}$ is
caused by the presence of $\Tau^W_0$. This opens a promising window for detecting
the large-scale tide. We set the redshift to be 0.5 to match most of the current
galaxy surveys, and the contribution from the long modes is smaller at high redshift,
assuming that the biases are unchanged, since both $\delta^W$ and $\Tau^W_m$ are
proportional to the linear growth factor.

\begin{figure}[t]
\centering
\includegraphics[width=0.495\textwidth]{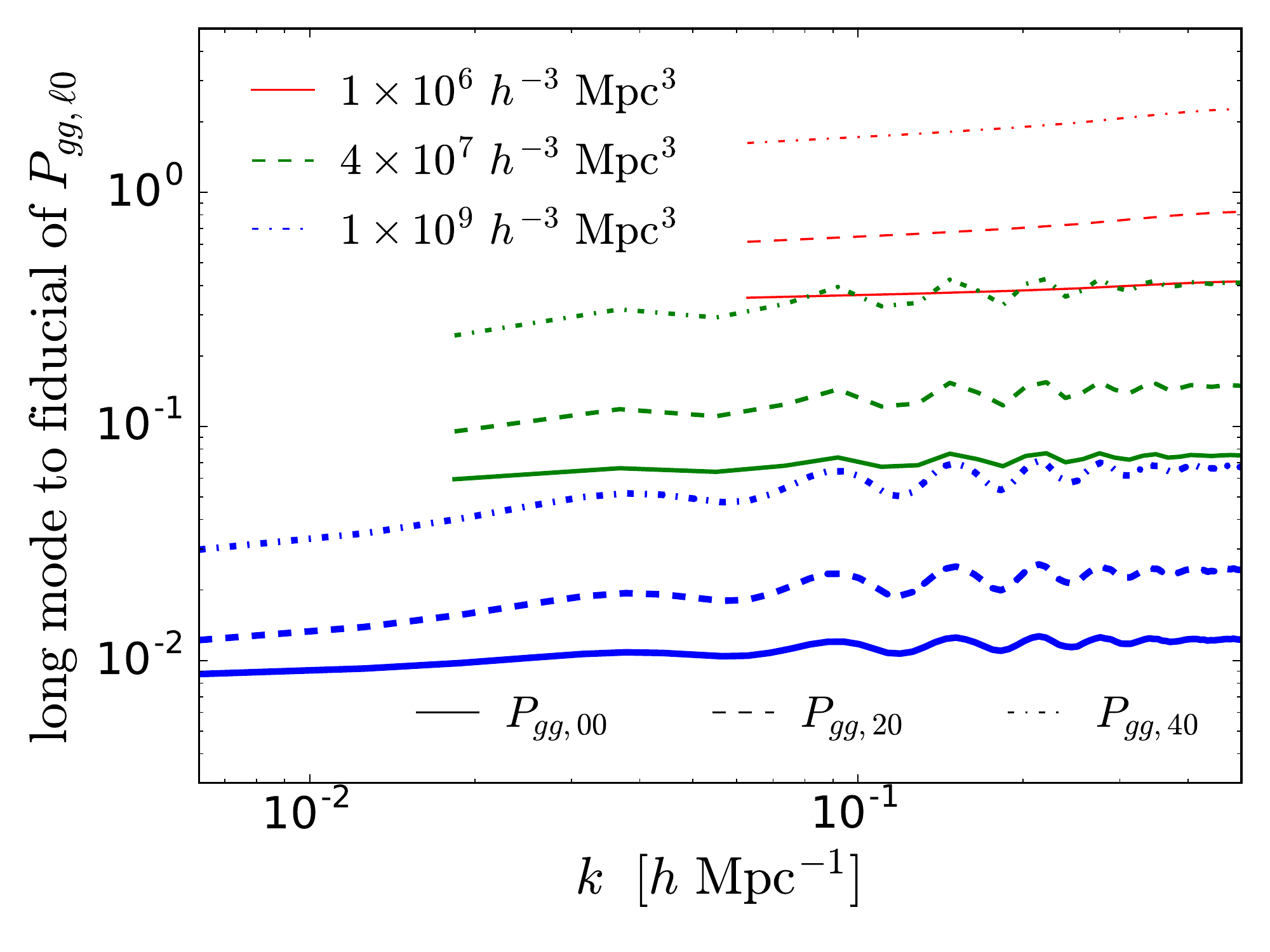}
\includegraphics[width=0.495\textwidth]{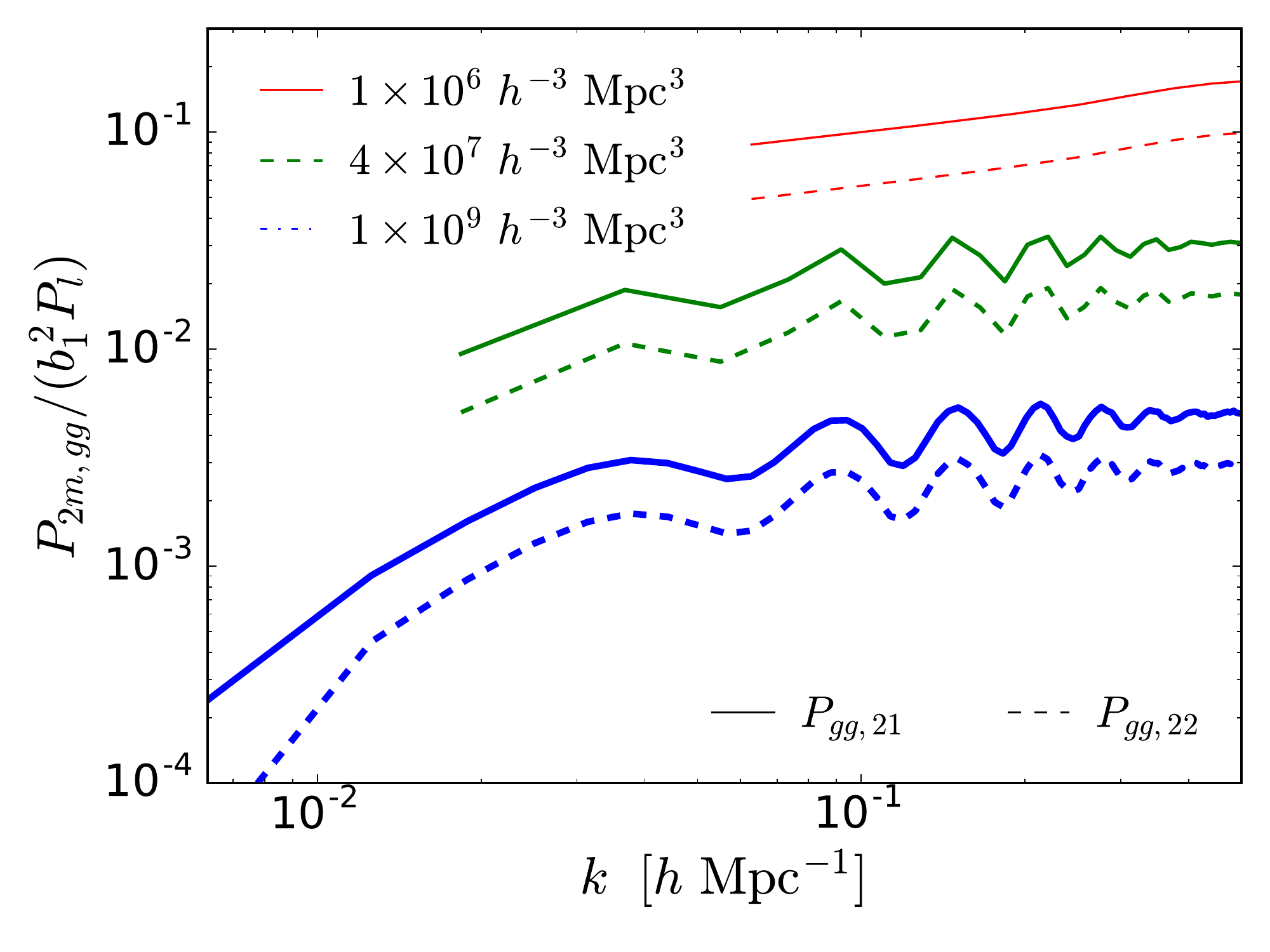}
\includegraphics[width=0.495\textwidth]{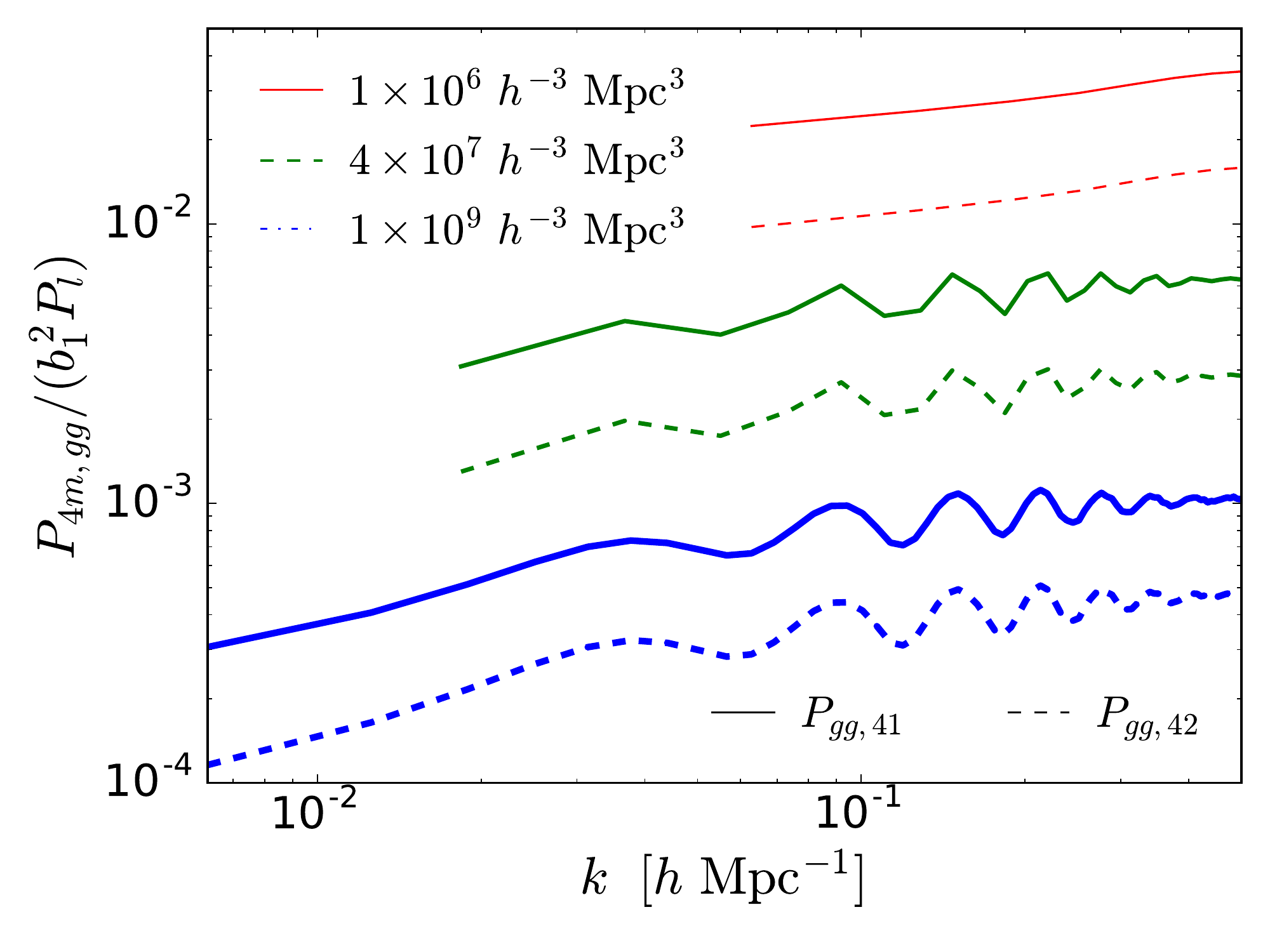}
\includegraphics[width=0.495\textwidth]{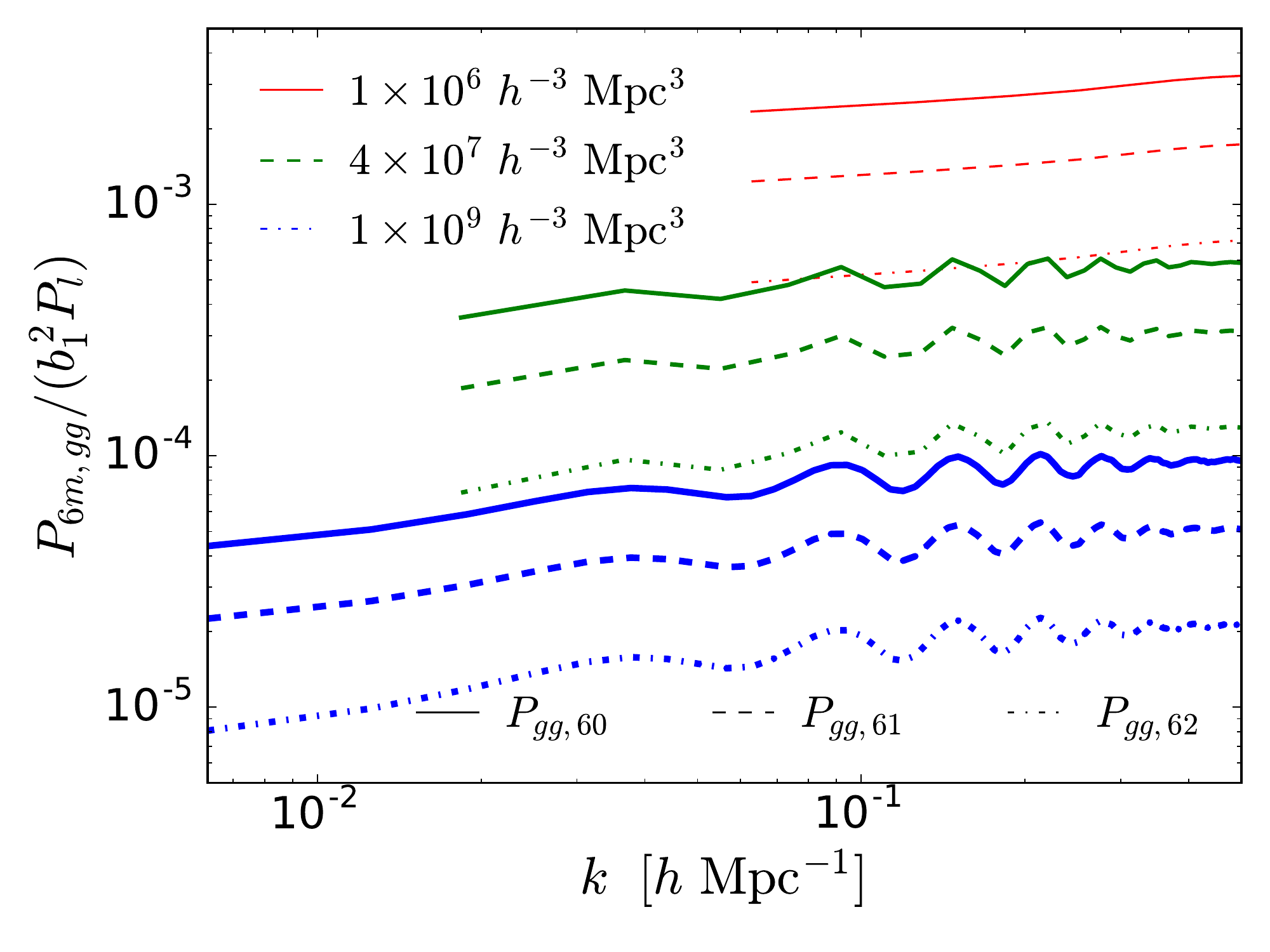}
\caption{(Top left) Ratio of the long mode contribution (terms associated with $\delta^W$
and $\Tau^W_0$) to fiducial ($\delta^W=\Tau^W_0=0$) $P_{gg,l0}$ for $\ell\le4$, i.e.
the Kaiser power spectrum, in redshift space at $z=0.5$. The solid, dashed, and dot-dashed
lines show $\ell=0$, $2$, and $4$, respectively. The other panels show relative sizes
to $b_1^2P_l$ of terms in redshift space at $z=0.5$ that are not sourced by the Kaiser
power spectrum.
(Top right) Ratio of $P_{gg,2m}$ to $b_1^2P_l$ for $m=1$ (solid) and 2 (dashed),
which are sourced respectively by $\Tau^W_1$ and $\Tau^W_2$.
(Bottom left) Same as the top right panel, but for $\ell=4$, i.e. $P_{gg,4m}$.
(Bottom right) Ratio of $P_{gg,6m}$ to $b_1^2P_l$ for $m=0$ (solid), 1 (dashed), and
2 (dot-dashed), which are sourced respectively by $\delta^W$ and $\Tau^W_0$, $\Tau^W_1$,
and $\Tau^W_2$. The red thin, green medium, and blue thick lines show volumes of $10^6$,
$4\times10^7$, and $10^9~\hMpcc$. The values of $\delta^W$ and $\Tau^W_m$ are set to be
their $1-\sigma$ expected values, i.e. $\sigma_{\delta^W}$ and $\sigma_{\Tau^W_m}$ with
spherical top-hat window function, respectively.
Note that the effect of using the local mean density to compute the power
spectrum, i.e. \refeq{local} is not included.}
\label{fig:Pggz}
\end{figure}

In \reffig{Pggz} we show the long mode contributions in redshift space at $z=0.5$.
The top left panel shows long mode contribution (terms associated with $\delta^W$
and $\Tau^W_0$) to fiducial ($\delta^W=\Tau^W_0=0$, i.e. the Kaiser power spectrum,
so independent of volume) $P_{gg,\ell0}$ for $\ell=0$ (solid), 2 (dashed), and 4
(dot-dashed). Interestingly, we find that for $\ell=4$ and $V=10^6~\hMpcc$ the long
mode contribution exceeds the fiducial Kaiser power spectrum. This indicates that
it is necessary to take the long mode contribution into account for the hexadecapole
of the galaxy redshift-space power spectrum when the survey volume is less than
$\sim10^6~\hMpcc$. The top right, bottom left, and bottom right panels show respectively
the ratios of $P_{gg,2m}$, $P_{gg,4m}$, and $P_{gg,6m}$ to $b_1^2P_l$. As in real
space, while these signals are smaller compared to the long mode contribution in
$P_{gg,\ell0}$, they are caused solely by the long modes ($\delta^W$ and $\Tau^W_0$
for $m=0$, $\Tau^W_1$ for $m=1$, and $\Tau^W_2$ for $m=2$), hence providing a great
potential to probe the large-scale perturbations. Unlike in real space, the signal
of the long modes depend both on the linear growth factor and the growth rate. Since
the growth factor and the growth rate have opposite redshift evolutions, the long
mode signal, assuming the biases do not evolve in redshift, does not have a clear
redshift evolution. 

Finally, we find that in general the effects of tides falls with $m$:
the relative impact of $\Tau^W_0$ is larger than that of $\Tau^W_1$
which is in turn larger than that of $\Tau^W_2$. This is true in both
real and redshift space.

In \reffig{Pggr} and \reffig{Pggz}, for clarity we do not include the
effect of using the local mean density to compute the power spectrum,
i.e. \refeq{local}. Since the corrections have the same angular dependencies
as the fiducial power spectrum, in real space the effect of the miscalibration
of the mean density only contributes to terms associated with $\delta^W$
in $P^r_{gg,00}$, and in redshift space to terms associated with $\delta^W$
as well as $\Tau^W_0$ in $P_{gg,\ell0}$ for $\ell=0$, 2, and 4.

\subsection{Estimator and covariance}
\label{sec:variance}
To measure $P_{gg,\ell m}(k)$ in a volume $V$, the simplest estimator is
\be
 \hP_{gg,\ell m}(k)=\frac{4\pi}{VN(k)}\sum_{k-\Delta k/2\le|\vk_i|\le k+\Delta k/2}
 \delta_g(\vk_i)\delta_g^*(\vk_i)Y^*_{\ell m}(\hk_i) \,,
\ee
where $N(k)$ is the number of independent Fourier modes. One can straightforwardly
show that this estimator is unbiased because in the continuous limit
\be
 \frac{1}{N(k)}\sum_{k-\Delta k/2\le|\vk_i|\le k+\Delta k/2}\to
 \frac{1}{4\pi}\int d^2\hk \,.
\ee

The covariance of the estimator can be computed as
\ba
 \:&{\rm cov}[\hP_{gg,\ell m}(k),\hP_{gg,\ell'm'}(k')] \vs
 =\:&\langle\hP_{gg,\ell m}(k)\hP_{gg,\ell'm'}(k')\rangle-\langle\hP_{gg,\ell m}(k)\rangle\langle\hP_{gg,\ell'm'}(k')\rangle \vs
 =\:&\frac{(4\pi)^2}{V^2N(k)N(k')}\sum_{ij}\langle\delta_g(\vk_i)\delta_g^*(\vk'_j)\rangle
 \langle\delta_g^*(\vk_i)\delta_g(\vk'_j)\rangle Y_{\ell m}^*(\hk_i)Y_{\ell'm'}^*(\hk'_j) \,,
\ea
where we assume that the covariance is dominated by the disconnected Gaussian
contribution and omit the full notation in the subscript of the summation. Note
that $\delta^W$ and $\Tau^W_m$ would also contribute to the covariance
\cite{Li:2014sga,Li:2014jra,Akitsu:2016leq,Akitsu:2017syq,Li:2017qgh}, but
we consider the survey to be large enough so that the super-sample covariance
is next-to-leading order correction. One can easily see that the covariance
is non-zero only if $\vk_i=\vk_j$, hence the covariance can be simplified to
\ba
 {\rm cov}[\hP_{gg,\ell m}(k),\hP_{gg,\ell'm'}(k)]\:&=
 \frac{(4\pi)^2}{N^2(k)}\sum_i[P_{gg}(\vk_i)+P_{\rm shot}]^2Y_{\ell m}^*(\hk_i)Y_{\ell'm'}^*(\hk_i) \vs
 \:&\to\frac{4\pi}{N(k)}\int d^2\hk[P_{gg}(\vk)+P_{\rm shot}]^2Y_{\ell m}^*(\hk)Y_{\ell'm'}^*(\hk) \,,
\ea
where $P_{\rm shot}$ is the shot noise. To proceed, we assume the galaxy
power spectrum is given by the Kaiser formalism, hence
\be
 {\rm cov}[\hP_{gg,\ell m}(k),\hP_{gg,\ell'm'}(k)]\approx
 \frac{4\pi}{N(k)}\int d^2\hk[(b_1+f\mu^2)^2P_l(k)+P_{\rm shot}]^2Y_{\ell m}^*(\hk)Y_{\ell'm'}^*(\hk) \,,
\label{eq:cov}
\ee
which is non-zero only if $m=m'$. We shall apply \refeq{cov} for the Fisher
analysis in \refsec{fisher}.

$P_{gg,\ell m}(k)$ is a complex number, and in practice we measure the real
and imaginary parts separately. Thus, the covariances are
\ba
 \:&{\rm cov}[\hP_{gg,\ell m}^R(k),\hP_{gg,\ell'm}^R(k)]=
 \frac{4\pi}{N_k}\int d^2\hk[(b_1+f\mu^2)^2P_l(k)+P_{\rm shot}]^2\Re[Y_{\ell m}^*(\hk)]\Re[Y_{\ell'm}^*(\hk)] \,, \vs
 \:&{\rm cov}[\hP_{gg,\ell m}^I(k),\hP_{gg,\ell'm}^I(k)]=
 \frac{4\pi}{N_k}\int d^2\hk[(b_1+f\mu^2)^2P_l(k)+P_{\rm shot}]^2\Im[Y_{\ell m}^*(\hk)]\Im[Y_{\ell'm}^*(\hk)] \,, \vs
 \:&{\rm cov}[\hP_{gg,\ell m}^R(k),\hP_{gg,\ell'm}^I(k)]=
 \frac{4\pi}{N_k}\int d^2\hk[(b_1+f\mu^2)^2P_l(k)+P_{\rm shot}]^2\Re[Y_{\ell m}^*(\hk)]\Im[Y_{\ell'm}^*(\hk)] \,,
\ea
and one can easily show that ${\rm cov}[\hP_{gg,\ell m}^R(k),\hP_{gg,\ell'm}^I(k)]=0$
for all possible $\ell,m$. Therefore, the only non-zero components are
\ba
 \:&{\rm cov}[\hP_{gg,\ell m}^R(k),\hP_{gg,\ell'm}^R(k)]=
 \frac{4\pi}{N(k)}\int d^2\hk[(b_1+f\mu^2)^2P_l(k)+P_{\rm shot}]^2\Re[Y_{\ell m}^*(\hk)]\Re[Y_{\ell'm}^*(\hk)] \,, \vs
 \:&{\rm cov}[\hP_{gg,\ell m}^I(k),\hP_{gg,\ell'm}^I(k)]=
 \frac{4\pi}{N(k)}\int d^2\hk[(b_1+f\mu^2)^2P_l(k)+P_{\rm shot}]^2\Im[Y_{\ell m}^*(\hk)]\Im[Y_{\ell'm}^*(\hk)] \,.
\ea
This means that for each $k$ the covariance matrix can be written as a
block-diagonal matrix, consisting covariances of $(\hP^R_{gg,00},\hP^R_{gg,20},\hP^R_{gg,40},\hP^R_{gg,60})$,
$(\hP^R_{gg,21},\hP^R_{gg,41},\hP^R_{gg,61})$, $(\hP^I_{gg,21},\hP^I_{gg,41},\hP^I_{gg,61})$,
$(\hP^R_{gg,22},\hP^R_{gg,42},\hP^R_{gg,62})$, and $(\hP^I_{gg,22},\hP^I_{gg,42},\hP^I_{gg,62})$.

\section{Fisher forecast}
\label{sec:fisher}
In the previous section we derive how galaxy power spectrum in a finite
volume would respond to the overdensity and tidal fields with wavelengths larger
than the volume. One specific example is the power spectrum in a galaxy redshift
survey, and it will be affected by the super-survey modes. This also means that
by measuring $P_{gg,\ell m}$ of this survey, it is possible to put constraints
on the super-survey overdensity and tidal fields, which are usually not directly
observable unless a large survey containing the current one is performed.

To explore the ability of measuring the long mode for a given survey, we apply
the Fisher matrix as
\be
 F_{\alpha\beta}=\sum_{k=k_{\rm min}}^{k_{\rm max}}\sum_{\ell\ell'}\sum_m
 \[{\rm cov}[P_{gg,\ell m}(k),P_{gg,\ell'm}(k)]\]^{-1}
 \frac{\partial P_{gg,\ell m}(k)}{\partial\theta_\alpha}\frac{\partial P_{gg,\ell'm}(k)}{\partial\theta_\beta} \,,
\ee
where $\theta_\alpha$ is the parameter of interest. The constraint on $\theta_\alpha$
as well as the correlation between $\theta_\alpha$ and $\theta_\beta$ are then
\be
 {\rm err}[\theta_\alpha]=\sqrt{\(F^{-1}\)_{\alpha\alpha}} \,, \quad
 {\rm corr}[\theta_\alpha,\theta_\beta]=\frac{\(F^{-1}\)_{\alpha\beta}}{{\rm err}[\theta_\alpha]{\rm err}[\theta_\beta]} \,.
\ee
For the fitting range we set $k_{\rm min}=k_F$ to be the fundamental frequency of
survey and explore the constraint for different $k_{\rm max}$. In this paper we
shall adopt the Planck cosmology \cite{Ade:2015xua}, i.e. $h=0.6803$, $\Omega_bh^2=0.0226$,
$\Omega_ch^2=0.1186$, $A_s=2.137\times10^{-9}$, and $n_s=0.9667$, hence the shape
of the power spectrum is fixed. We fix the redshift to be 0.5 because it is the
redshift at which most galaxy surveys are performed, but the results can be straightforwardly
generalized to other redshifts.
The parameters of interest are
$\theta_\alpha\in(b_1,b_2,b_{s^2},f,\delta^W,\Tau^W_0,\Tau^{W,R}_1,\Tau^{W,I}_1,\Tau^{W,R}_2,\Tau^{W,I}_2)$,
where $\Tau^W_1$ and $\Tau^W_2$ are complex numbers so there are four parameters in
total that one can measure. We set the fiducial values of the biases and growth rate
to be $b_1=2$, $b_2=0.3$, $b_{s^2}=-\frac{4}{7}(b_1-1)=-0.57$, and $f(z=0.5)=0.75$,
and for the long mode we set the fiducial value to be the $1-\sigma$ expected value
for the corresponding volume, assuming a spherical top-hat window function. In the
following we shall separately discuss the results in real and redshift space, and
in real space we set $f=0$ so the number of parameters is nine.

\subsection{Real space}
\label{sec:rspace}
Let us begin with the Fisher analysis in real space, in which the power spectra are
given in \refeq{Pggr}. Moreover, since the global mean density is unknown if only
a finite survey is performed, only the local mean density can be used to measure the
power spectrum, hence there is an additional contribution to the response from the
miscalibration of the mean density. We thus use \refeq{local} to mimic this effect,
and only $P^r_{gg,00}$ contains the additional contribution.

We first notice that Fisher matrix is not positive definite. This happens because
there are nine parameters to be determined, but from \refeq{Pggr} one can only
measure eight scale dependencies: two from $P^r_{gg,00}$, two from $P^r_{gg,20}$,
and two from $P^r_{gg,21}$ and $P^r_{gg,22}$ respectively because they are complex
numbers and have the same scale dependence as $P^r_{gg,20}$ hence only four independent
amplitudes can be measured. Since the main focus of this paper is to probe the long
mode as well as to study their impact on $b_1$ and $f$, we shall include priors of
$\pm1$ on $b_2$ and $b_{s^2}$. These priors are sufficiently strong that they break
the perfect degeneracy to the extent that more constraining prior has negligible
effects on the results. 

\begin{figure}[t]
\centering
\includegraphics[width=0.495\textwidth]{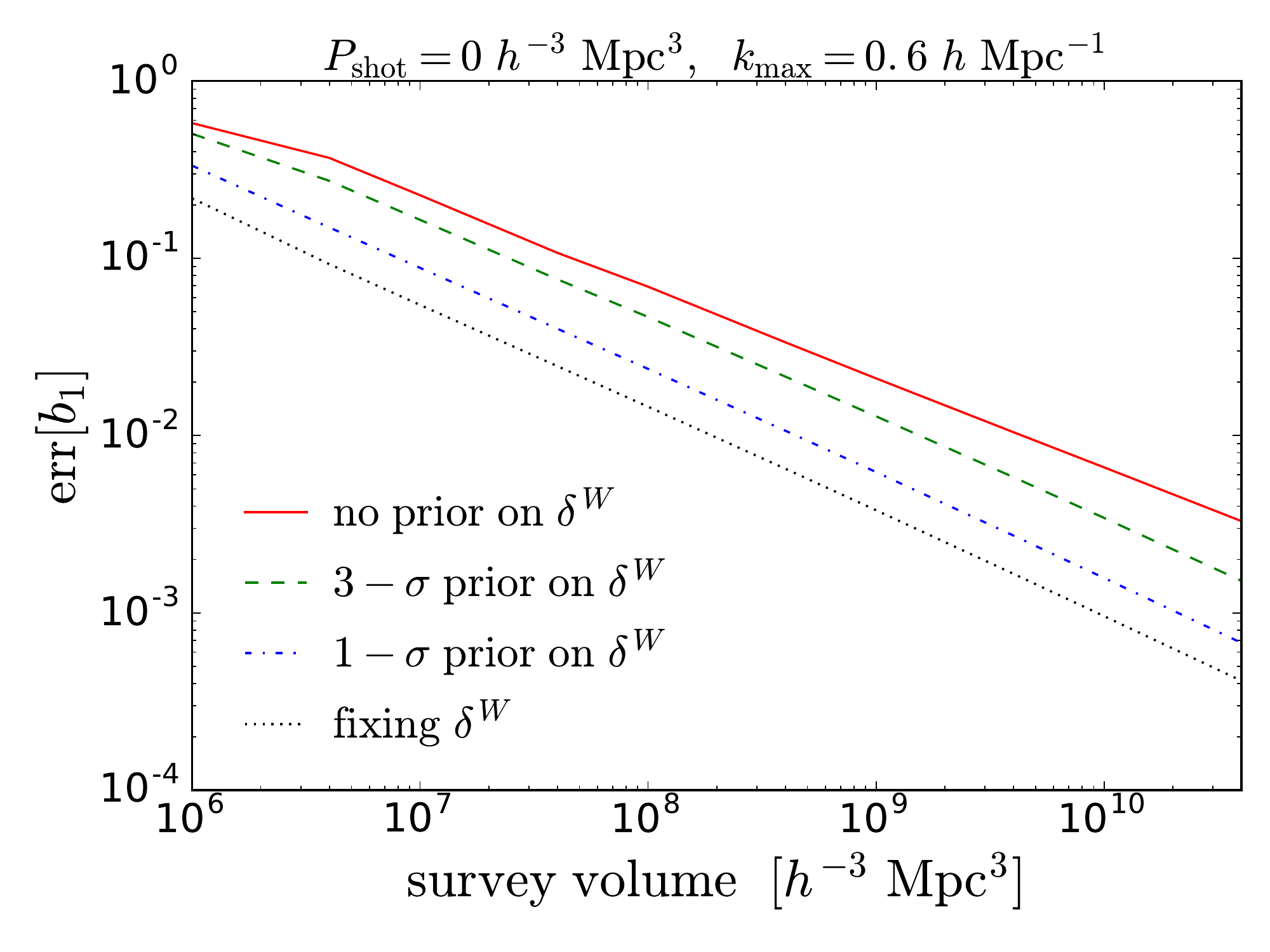}
\includegraphics[width=0.495\textwidth]{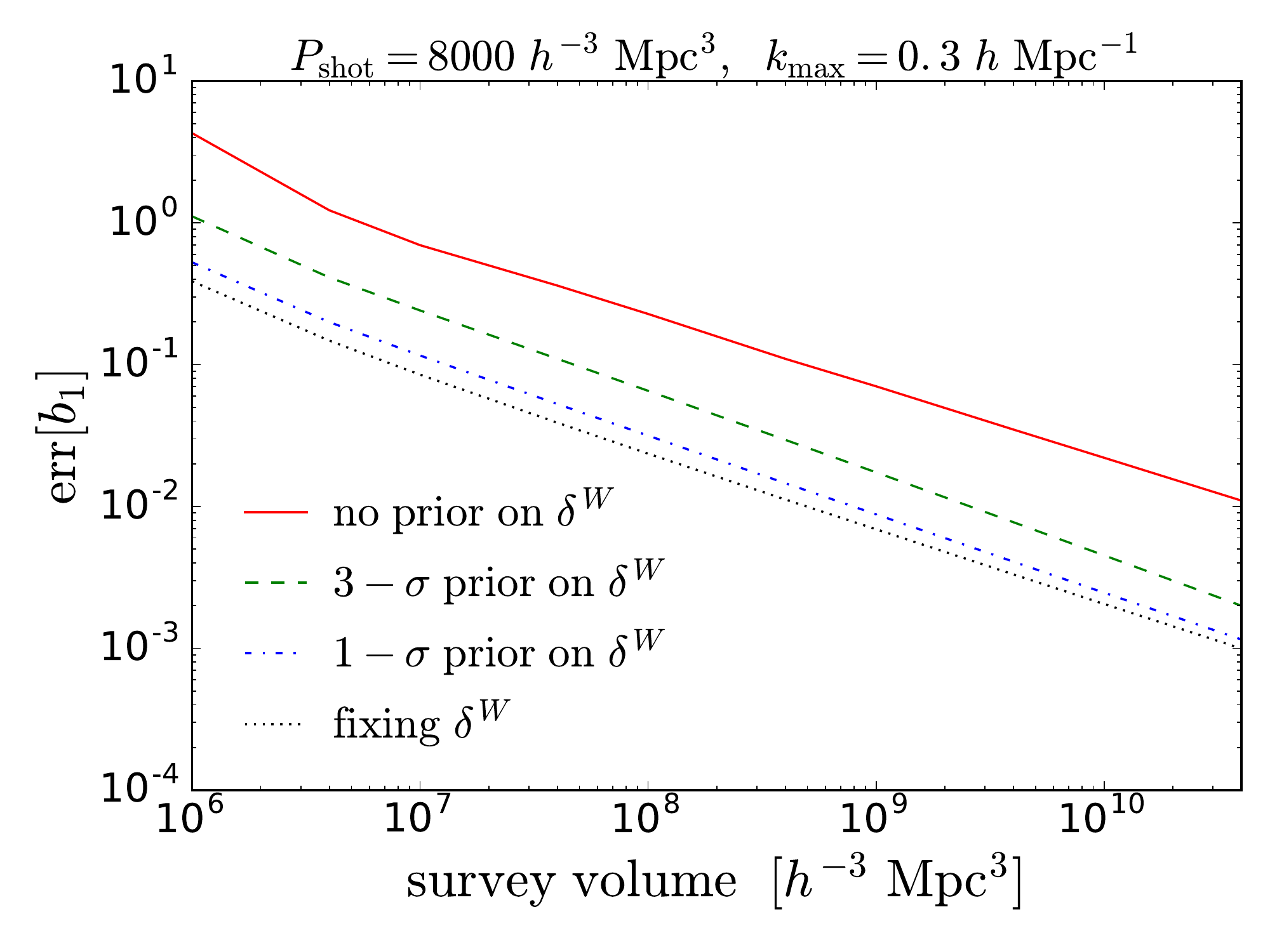}
\caption{$1-\sigma$ constraint on $b_1$ from the real-space galaxy power spectrum
as a function of survey volume for two $P_{\rm shot}$ and $k_{\rm max}$. The left
panel shows a cosmic variance limited survey with high $k_{\rm max}$, whereas the
right panel shows a BOSS-like survey number density with a realistic $k_{\rm max}$.
Lines with different colors and styles show various priors on $\delta^W$. Note
that a large number of $\sigma$ is for a weaker prior.}
\label{fig:sigmab1_rspace}
\end{figure}

Adding priors to $b_2$ and $b_{s^2}$, the Fisher matrix becomes invertible,
and we find that $b_1$ and $\delta^W$ are highly correlated. Specifically,
for $10^6\le V/(\hMpcc)\le4\times10^{10}$, $0\le P_{\rm shot}/(\hMpcc)\le8000$
(cosmic variance limited to the BOSS-like survey number density \cite{Alam:2016hwk}),
and $0.1\le k_{\rm max}/(\ihMpc)\le0.5$, the correlation between $b_1$ and
$\delta^W$ is greater than 0.95. The large correlation is not surprising as
the ratio shown in the left panel of \reffig{Pggr} is quite scale independent,
and the correction from the miscalibration of the mean density has identical
scale dependence as the fiducial power spectrum. This means that the constraint
on $b_1$ will be largely determined by the knowledge on $\delta^W$. \refFig{sigmab1_rspace}
shows the $1-\sigma$ constraint on $b_1$ as a function of survey volume for
two $P_{\rm shot}$ and $k_{\rm max}$: the left panel shows no shot noise whereas
the right panel shows a BOSS-like survey number density. It is evident that
the constraint on $b_1$ is largely dominated by the prior on $\delta^W$.
Specifically, adding a $1-\sigma$ prior on $\delta^W$ can improve the constraint
on $b_1$ by an order of magnitude compared to no prior. The main caveat in
\reffig{sigmab1_rspace} is that we only consider the leading-order galaxy
power spectrum, and in reality the small-scale nonlinearities in matter power
spectrum and galaxy bias reduce the information on linear bias. Nevertheless,
\reffig{sigmab1_rspace} clearly demonstrates the impact from the knowledge of
$\delta^W$ on $b_1$ constraint.

\begin{figure}[t]
\centering
\includegraphics[width=0.495\textwidth]{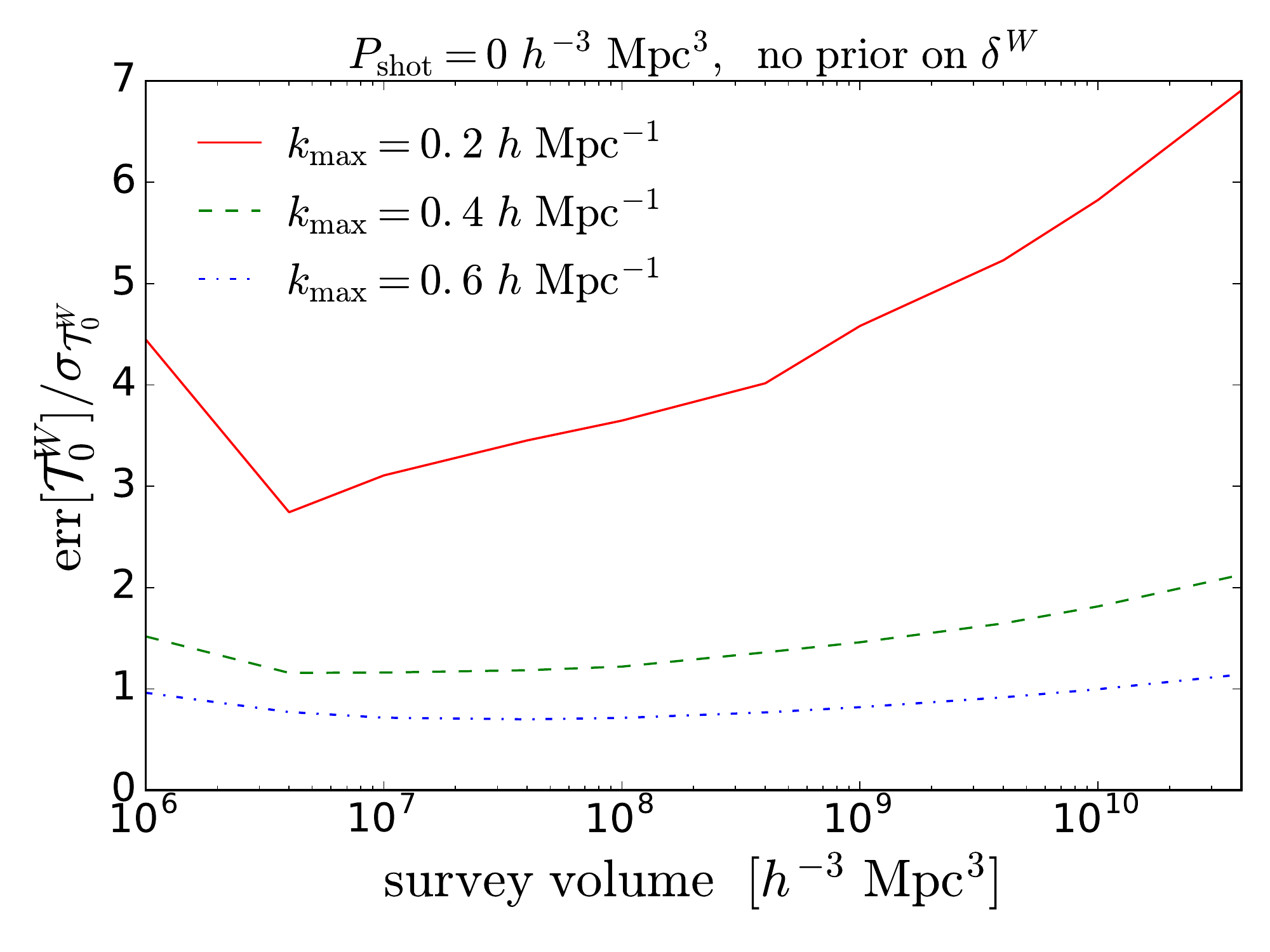}
\includegraphics[width=0.495\textwidth]{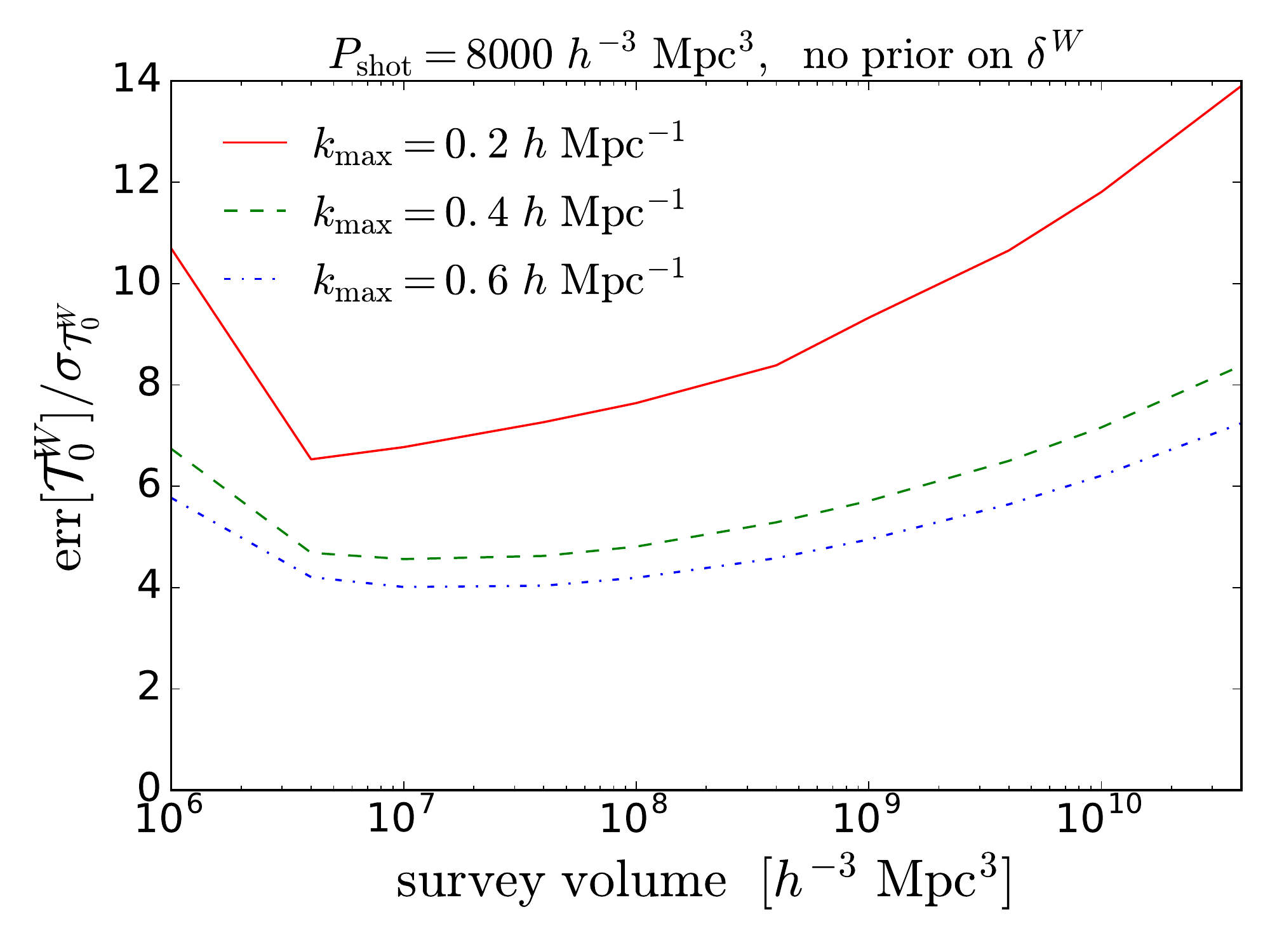}
\caption{Ratio of the $1-\sigma$ constraint on $\Tau^W_0$ to its expected value,
i.e. ${\rm err}[\Tau^W_0]/\sigma_{\Tau^W_0}$, as a function of survey volume for
$P_{\rm shot}=0$ (left) and $8000~\hMpcc$ (right). The red solid, green dashed,
and blue dot-dashed lines show respectively $k_{\rm max}=0.2$, 0.4, and $0.6~\ihMpc$,
whereas the black dotted line is for $\sigma_{\Tau^W_0}$. We do not include any
prior on $\delta^W$, as it has negligible impact on the constraint on $\Tau^W_0$.
Furthermore, we only present the constraint on $\Tau^W_0$ because the results are
identical for $\Tau^{W,R}_1$, $\Tau^{W,I}_1$, $\Tau^{W,R}_2$, and $\Tau^{W,I}_2$.}
\label{fig:sigmaT0_rspace}
\end{figure}

For the large-scale tidal fields, we find that the absolute value of the correlation
between $b_1$ and $\Tau^W_m$ is less than 0.1 if the survey volume is greater than
$10^8~\hMpcc$. The reason is that the constraint on $b_1$ is mainly from $P^r_{gg,00}$,
for which $\Tau^W_m$ do not contribute. As a result, the prior on $\delta^W$ has
negligible effect on the constraints on $\Tau^W_m$, meaning that $\Tau^W_m$ can be
measured robustly in real space. \refFig{sigmaT0_rspace} shows the ratio of the
$1-\sigma$ constraint on $\Tau^W_0$ to its expected value, i.e. ${\rm err}[\Tau^W_0]/\sigma_{\Tau^W_0}$,
as a function of survey volume for $P_{\rm shot}=0$ (left) and $8000~\hMpcc$ (right).
We only present the constraint on $\Tau^W_0$ since the results are identical for
$\Tau^{W,R}_1$, $\Tau^{W,I}_1$, $\Tau^{W,R}_2$, and $\Tau^{W,I}_2$. The red solid,
green dashed, and blue dot-dashed lines show respectively $k_{\rm max}=0.2$, 0.4,
and $0.6~\ihMpc$. We find that the constraint depends significantly on the shot noise.
Specifically, for the cosmic variance limited survey ($P_{\rm shot}=0$) $\sigma_{\Tau^W_0}$
can be achieved if $k_{\rm max}=0.6~\ihMpc$, while for BOSS-like survey number density
the constraint on $\sigma_{\Tau^W_0}$ worsen by more than a factor of two. Note that
while it may seem unrealistic to adopt $k_{\rm max}=0.6~\ihMpc$ for a galaxy redshift
survey, the presence of $P^r_{gg,2m}$ cannot be produced by nonlinear evolution and
is a distinct feature of large-scale tidal fields. Therefore, in the spirit of putting
an upper limit on $\Tau^W_m$, it is justified to use much higher wavenumber. Moreover,
on small scales the survey window function tends to be isotropized \cite{Sato:2013hea},
hence the contamination on $P^r_{gg,2m}$ due to the survey window function becomes
less important \cite{Sugiyama:2017ggb}. In \reffig{sigmaT0_rspace} we also notice
a minimum ratio at $\sim3\times10^6~\hMpcc$. This is because for small volume the
signal due to $\Tau^W_m$ is larger and for large volume there are more modes one
can access to constrain $\Tau^W_m$. Hence there is a sweet spot in the survey volume
for constraining the large-scale tidal fields, and the exact value depends on $P_{\rm shot}$
and $k_{\rm max}$.

\subsection{Redshift space}
\label{sec:zspace}
Let us now turn to the Fisher analysis in redshift space, in which the power
spectra are given in \refeqs{Pggm0}{Pggm2} with ten parameters. As in real
space, we adopt \refeq{local} to account for the miscalibration of the mean
density when measuring the power spectrum in a finite survey, so $P_{gg,00}$,
$P_{gg,20}$, and $P_{gg,40}$ receive additional contribution. To make the
inversion of the Fisher matrix stable, we also include a prior of $\pm1$ on
$b_2$ and $b_{s^2}$. Since the conclusions are insensitive to the choice of
the survey parameters, in this section we shall fix $P_{\rm shot}=8000~\hMpcc$
and $k_{\rm max}=0.3~\ihMpc$ for a more realistic forecast.

\begin{figure}[t]
\centering
\includegraphics[width=0.495\textwidth]{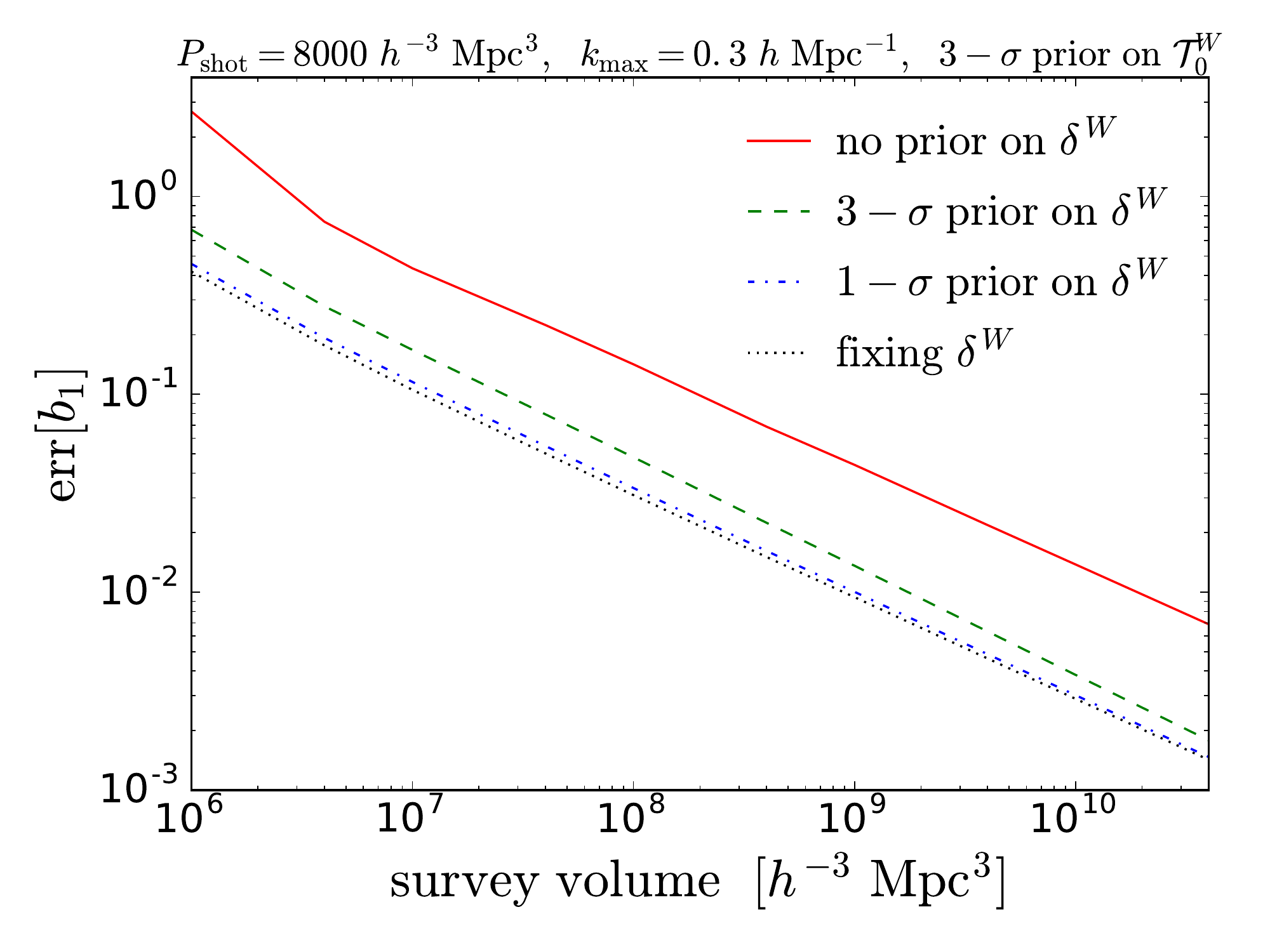}
\includegraphics[width=0.495\textwidth]{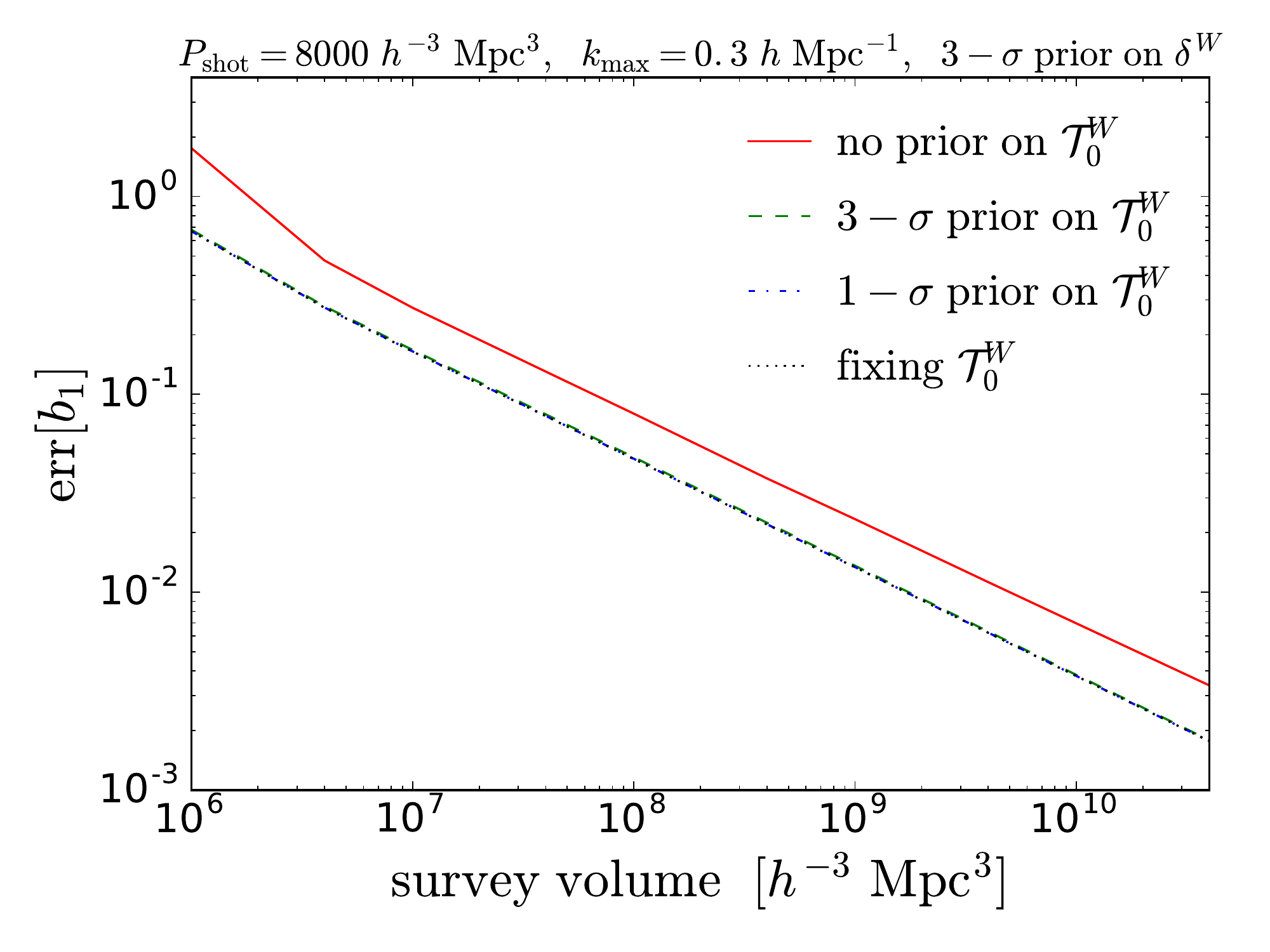}
\caption{(Left) $1-\sigma$ constraint on $b_1$ from the redshift-space galaxy
power spectrum as a function of survey volume for $P_{\rm shot}=8000~\hMpcc$,
$k_{\rm max}=0.3~\ihMpc$, and a $3-\sigma$ prior on $\Tau^W_0$. Lines with
different colors and styles show various priors on $\delta^W$. (Right) Same
as the left panel, but with a $3-\sigma$ prior on $\delta^W$. Lines with
different colors and styles show various priors on $\Tau^W_0$.}
\label{fig:sigmab1_zspace}
\end{figure}

We first examine the constraint on $b_1$, which is dominated by $P_{gg,00}$,
$P_{gg,20}$, and $P_{gg,40}$ due to their large signal-to-noise ratios compared
to the rest of $P_{gg,\ell m}$. Since only $\delta^W$ and $\Tau^W_0$ contribute
to $m=0$ components, we expect that $b_1$ is mostly degenerate with them. As in
real space, we find that $b_1$ and $\delta^W$ are highly correlated regardless
of the survey parameters and the priors on $\Tau^W_m$, with correlation coefficients
greater than 0.8, hence the constraint on $b_1$ will depend strongly on the prior
on $\delta^W$. The left panel of \reffig{sigmab1_zspace} shows the $1-\sigma$
constraint on $b_1$ from the redshift-space galaxy power spectrum with a $3-\sigma$
prior on $\Tau^W_0$ as a function of survey volume. We find that different priors
on $\delta^W$ can affect the constraint on $b_1$ by more than an order of magnitude,
and this finding is consistent as in real space. For the correlation coefficient
between $b_1$ and $\Tau^W_0$, we find it to be less than -0.5 when no prior on
$\delta^W$ is included. The anti-correlation increases when we include a prior
on $\delta^W$, hence we expect to see some dependence on the prior on $\Tau^W_0$
of the constraint on $b_1$. The right panel of \reffig{sigmab1_zspace} shows the
$1-\sigma$ constraint on $b_1$ with a $3-\sigma$ prior on $\delta^W$. We find that
as long as there is some prior on $\Tau^W_0$, the constraint on $b_1$ converges
well, indicating that a reliable constraint on $b_1$ can be obtained even with
a conservative ($3-\sigma$) prior on $\Tau^W_0$. Interestingly, we notice that
if there is no prior on $\delta^W$, then the prior on $\Tau^W_0$ has negligible
effect on the constraint on $b_1$. This further reinforces the strong correlation
between $b_1$ and $\delta^W$.

\begin{figure}[t]
\centering
\includegraphics[width=0.495\textwidth]{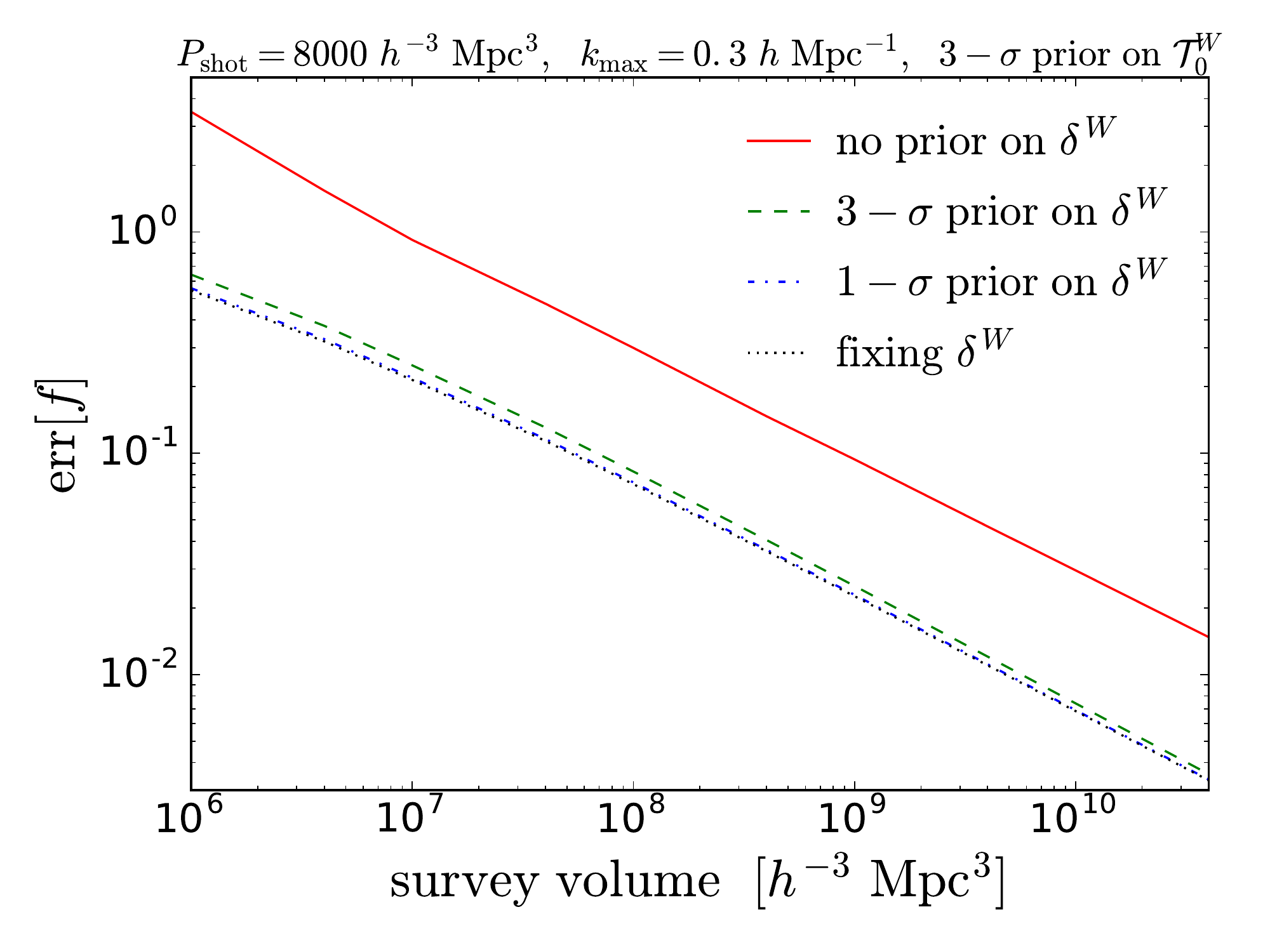}
\includegraphics[width=0.495\textwidth]{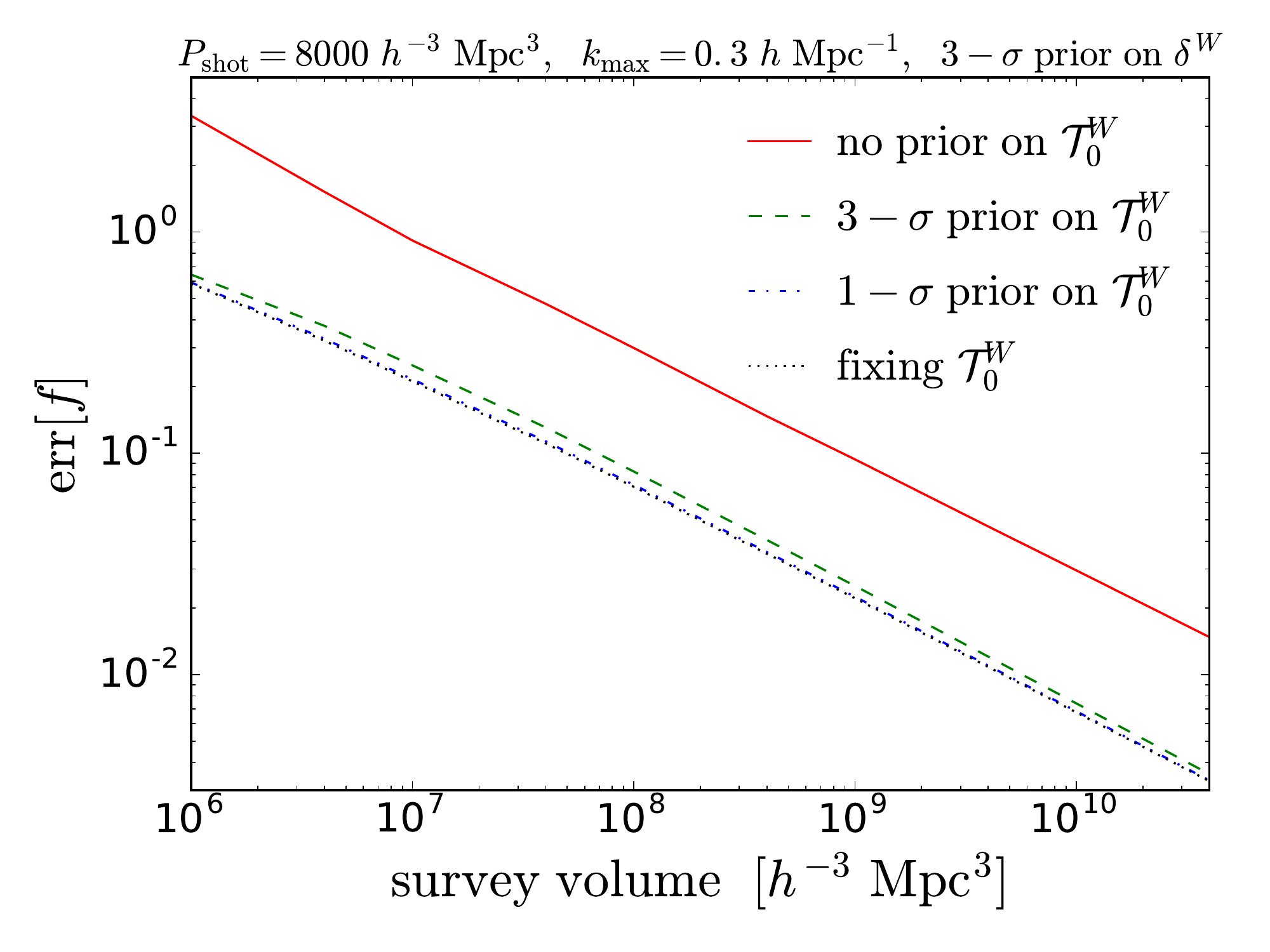}
\caption{(Left) 1-$\sigma$ constraint on $f$ from the redshift-space galaxy
power spectrum as a function of survey volume for $P_{\rm shot}=8000~\hMpcc$,
$k_{\rm max}=0.3~\ihMpc$, and a $3-\sigma$ prior on $\Tau^W_0$. Lines with
different colors and styles show various priors on $\delta^W$. (Right) Same
as the left panel, but with a $3-\sigma$ prior on $\delta^W$. Lines with
different colors and styles show various priors on $\Tau^W_0$.}
\label{fig:sigmaf_zspace}
\end{figure}

We next examine the constraint on $f$, which as $b_1$ is dominated by $P_{gg,00}$,
$P_{gg,20}$, and $P_{gg,40}$, so we focus on the degeneracy with $\delta^W$ and
$\Tau^W_0$ as well. We find that the absolute value of the correlation coefficient
between $f$ and $\delta^W$ is less than 0.35 (weaker correlation for larger volume)
and has almost no dependence on the prior on $\Tau^W_0$, whereas the correlation
coefficient between $f$ and $\Tau^W_0$ changes from $\sim-0.85$ for no prior on
$\delta^W$ to less than -0.9 for a conservative $3-\sigma$ prior on $\delta^W$.
The strong effect suggests that the constraint on $f$ will be dependent on both
the priors on $\delta^W$ and $\Tau^W_0$, and indeed we find that just adding one
prior on either $\delta^W$ or $\Tau^W_0$ does not improve the constraint on $f$.
\refFig{sigmaf_zspace} shows the constraint on $f$ from the redshift-space galaxy
power spectrum with various priors on $\delta^W$ and $\Tau^W_0$ as a function of
the survey volume. Including conservative $3-\sigma$ priors on both $\delta^W$
and $\Tau^W_0$ reduces the constraint on $f$ significantly compared to the one with
only one prior on either $\delta^W$ or $\Tau^W_0$, and the result converges well
with that of fixing both $\delta^W$ and $\Tau^W_0$. This implies that for future
large-scale structure analysis it is sufficient to obtain a rigorous constraint
on $f$ as long as $3-\sigma$ priors on both $\delta^W$ and $\Tau^W$ are added.

Since $b_1$ and $f$ are mostly degenerate with $\delta^W$ and $\Tau^W_0$, it is
natural to ask whether the inclusion of $P_{gg,60}$, which can only be produced
by $\delta^W$ and $\Tau^W_0$, improves the constraint on $b_1$ and $f$ or not.
To address this question, we perform the Fisher analysis using the observables
of $(P_{gg,00},P_{gg,20},P_{gg,40})$ and $(P_{gg,00},P_{gg,20},P_{gg,40},P_{gg,60})$.
However, we find that the constraint on $b_1$ and $f$ is insensitive to the
presence of $P_{gg,60}$. This is likely due to the low signal-to-noise ratio of
$P_{gg,60}$, because even if $\delta^W$ ($\Tau^W_0$) is fixed, the constraint
on $\Tau^W_0$ ($\delta^W$) reduces only by a few percent regardless of the
existence of $P_{gg,60}$. Therefore, it is better to include priors on $\delta^W$
and $\Tau^W_0$ for acquiring reliable constraints on both $b_1$ and $f$.

\begin{figure}[t]
\centering
\includegraphics[width=0.495\textwidth]{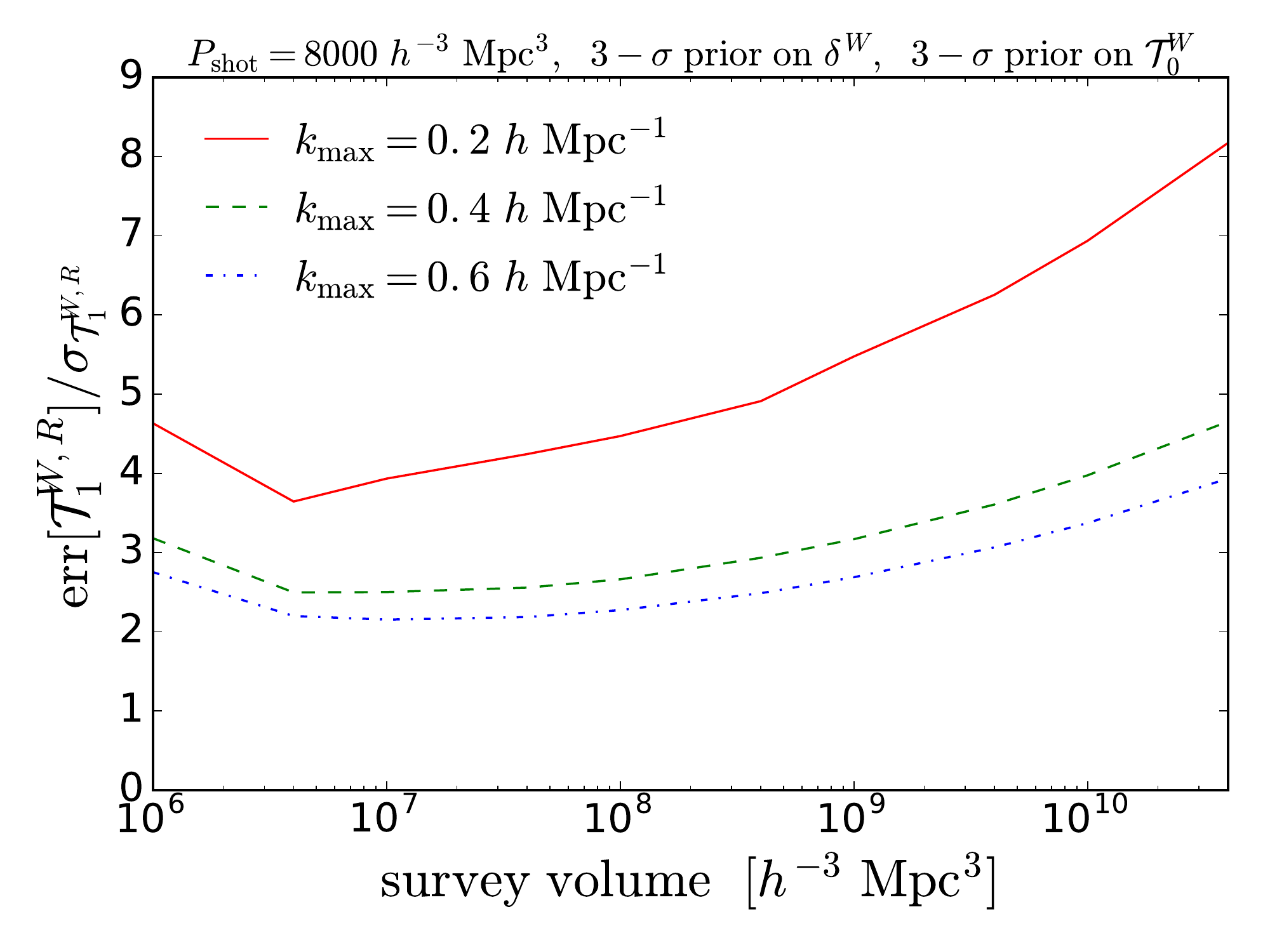}
\includegraphics[width=0.495\textwidth]{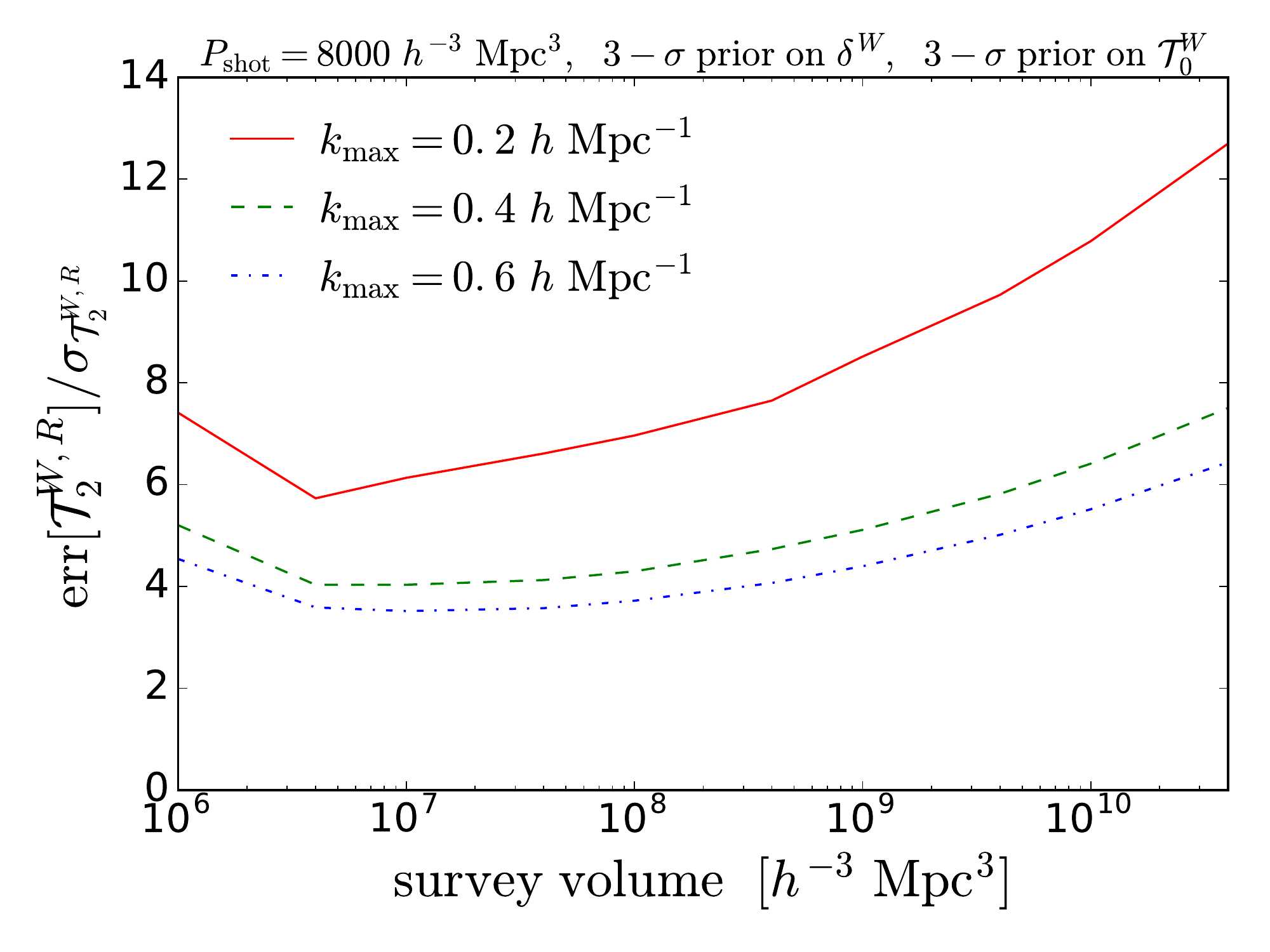}
\caption{(Left) Ratio of the 1-$\sigma$ constraint from the galaxy redshift-space
power spectrum on $\Tau^{W,R}_1$ to its expected value, i.e.
${\rm err}[\Tau^{W,R}_1]/\sigma_{\Tau^{W,R}_1}$, as a function of survey volume
for $P_{\rm shot}=8000~\hMpcc$, and $3-\sigma$ priors on $\delta^W$ and $\Tau^W_0$.
The constraint on $\Tau^{W,I}_1$ is identical to $\Tau^{W,R}_1$. Lines with different
colors and styles show various $k_{\rm max}$. (Right) Same as the left panel, but
for the $1-\sigma$ constraint on $\Tau^{W,R}_2$.}
\label{fig:sigmaT1T2_zspace}
\end{figure}

\begin{figure}[t]
\centering
\includegraphics[width=0.495\textwidth]{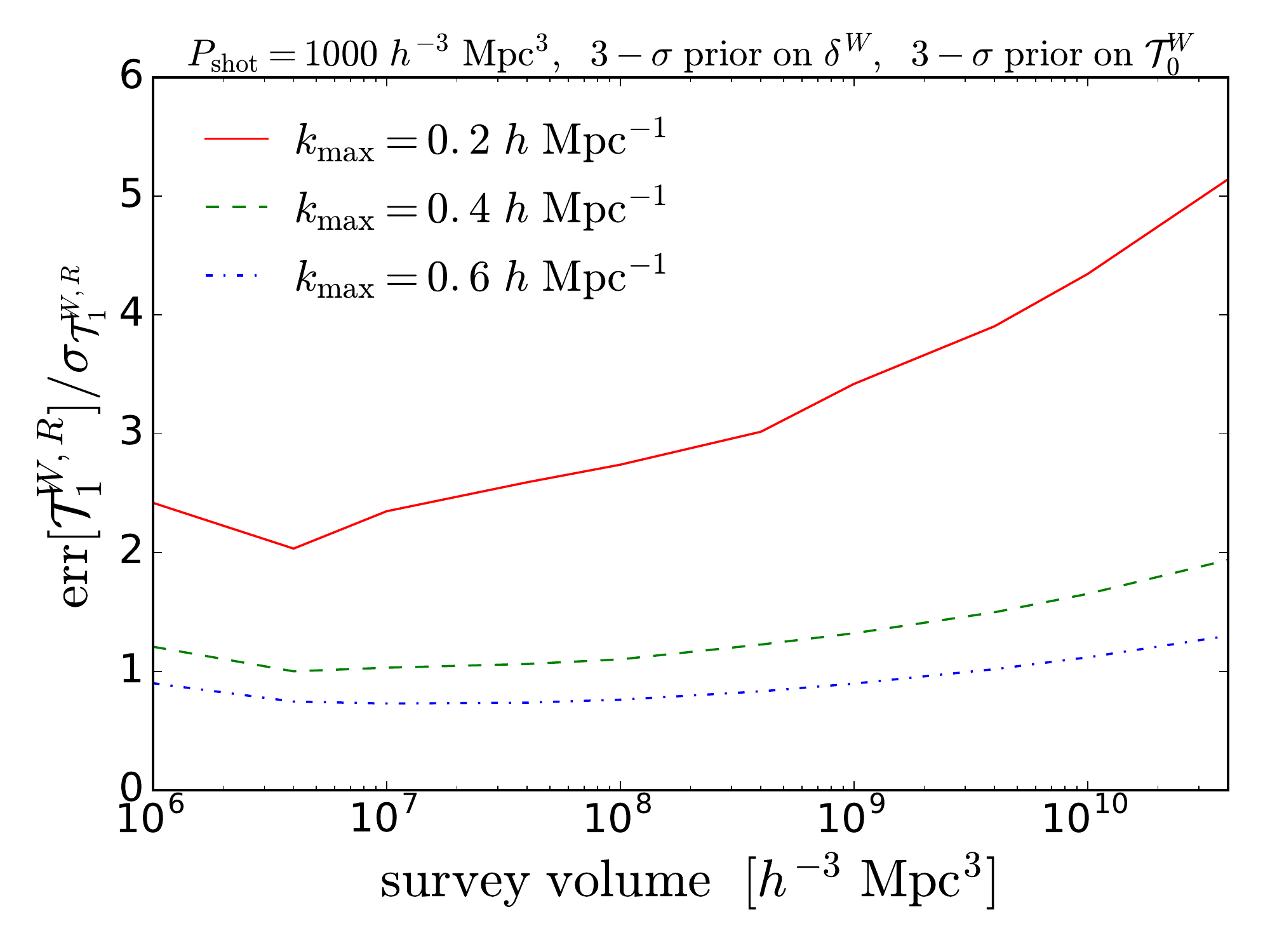}
\includegraphics[width=0.495\textwidth]{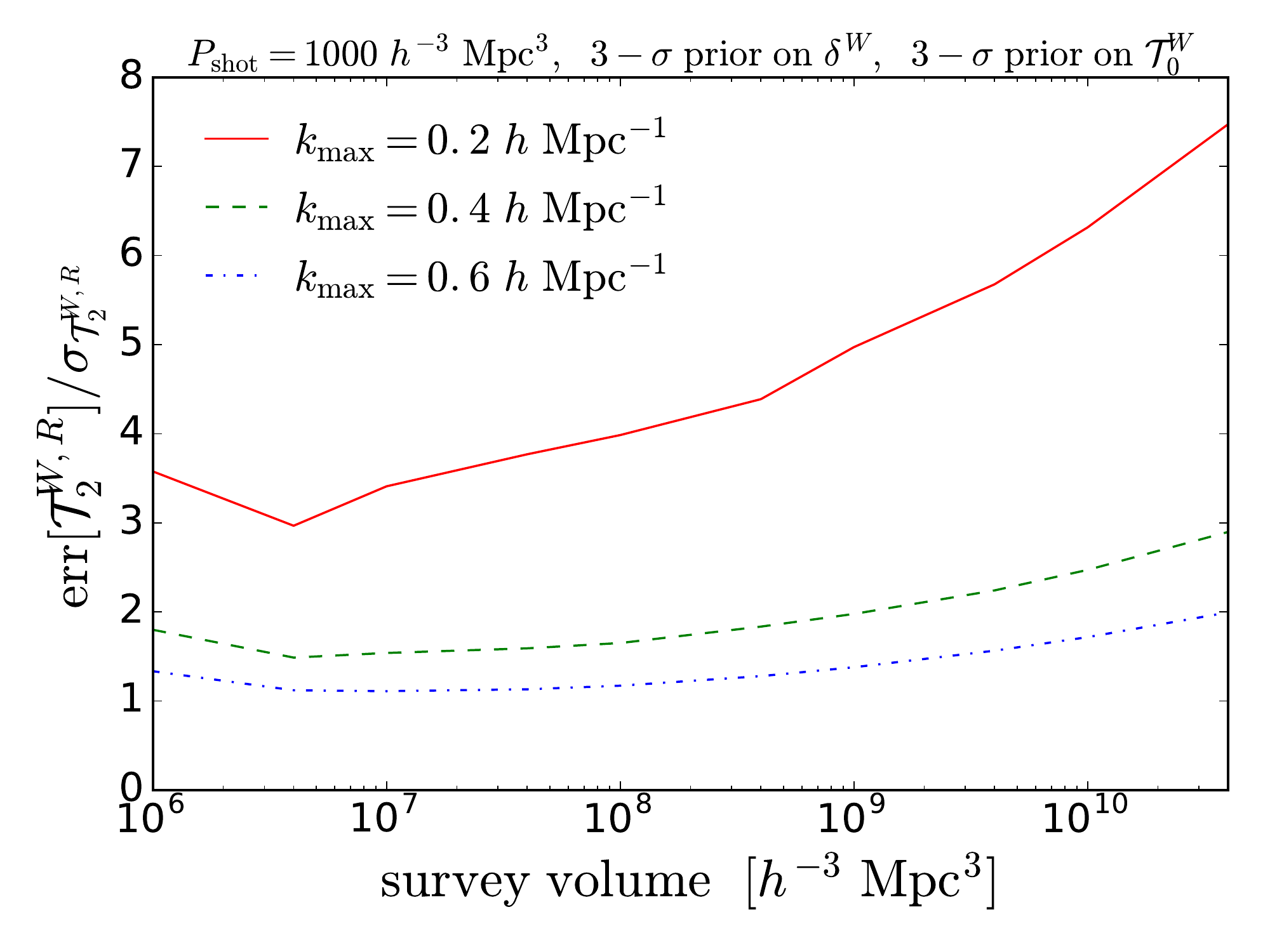}
\caption{Same as \reffig{sigmaT1T2_zspace}, but with $P_{\rm shot}=1000~\hMpcc$.}
\label{fig:sigmaT1T2_zspace_lowPshot}
\end{figure}

While it is difficult to measure $\delta^W$ and $\Tau^W_0$ due to their low
signal-to-noise, $\Tau^W_1$ and $\Tau^W_2$ can be probed because they are the
only sources that can contribute respectively to $P_{gg,\ell1}$ and $P_{gg,\ell2}$.
Moreover, we find that for a survey volume greater than $10^8~\hMpcc$, the
absolute values of the correlation coefficients between $b_1$ and $\Tau^W_m$
as well as $f$ and $\Tau^W_m$ are less than 0.1 for $m\ge1$. \refFig{sigmaT1T2_zspace}
shows the ratios of the $1-\sigma$ constraints from the redshift-space galaxy
power spectrum on $\Tau^{W,R}_1$ (left) and $\Tau^{W,R}_2$ (right) to their
expected values, i.e. ${\rm err}[\Tau^{W,R}_1]/\sigma_{\Tau^{W,R}_1}$ and
${\rm err}[\Tau^{W,R}_2]/\sigma_{\Tau^{W,R}_2}$, as a function of survey volume
for $P_{\rm shot}=8000~\hMpcc$, and $3-\sigma$ priors on $\delta^W$ and $\Tau^W_0$.
The constraints on the imaginary part are identical to the real part. We find
that although it is possible to put upper bounds on the super-survey tidal fields,
the constraints are quite weak, ranging from $\sim3$ to 4 times of the $\Lambda$CDM
expected values for a survey of $10^9~\hMpcc$, even if a large $k_{\rm max}$ of
$0.6~\ihMpc$ is used. The weak constraints are mainly driven by the large shot
noise, and for a high number density survey with $P_{\rm shot}=1000~\hMpcc$ as
shown in \reffig{sigmaT1T2_zspace_lowPshot}, one can in general put $\sim1-\sigma$
constraints on the super-survey tidal fields as long as a high $k_{\rm max}$ is
adopted. This is consistent with the finding in Ref.~\cite{Akitsu:2017syq} when
all cosmological parameters are fixed, though they only forecasted the constraint
for $\Tau^W_0$.

\section{Conclusion}
\label{sec:conclusion}
We have generalized the Kaiser formula in a finite volume with
large-scale overdensity and tidal fields. We find that $\ell=0,2,4,6$
and $m=0,1,2$ modes are present in spherical harmonic expansion.
The linear power spectrum and mean overdensity generate the azimuthally
symmetric $\ell=0,2,4,6$, $m=0$ modes of the power spectrum. The
tidal fields, written as a quadrupole, generate $\ell=2,4,6$ modes
of the power spectrum, separately for each $m$ components. Hence,
the azimuthal symmetry of the Kaiser power spectrum in a finite
volume is broken due to the presence of the large-scale tidal fields.

The first non-trivial result of writing equations in this way is that
we note that the tidal contribution to the azimuthally symmetric power
spectrum is sourced by just one of the five degrees of freedom present
in the tidal fields. This allows a more natural way of marginalizing
over this uncertainty. Our numerical calculation shows that the relative
size of effect associated with the tidal fields decrease with increasing
$m$.

The additional small-scale physics cannot break the basic symmetries
of the problem. Hence, beyond second order physics that is still linear
in $\Tau^W_m$ will only affect the same $m$ modes, although it can in
principle affect arbitrarily large $\ell$, for example, fingers of God.
However, terms that are quadratic in $\Tau^W_m$ will in general couple to
the sum and difference of two $m$ components.

As a concrete example, we made Fisher forecasts for a galaxy redshift
survey in determining the galaxy bias parameters, the growth rate $f$,
and the super-survey overdensity $\delta^W$ and tidal fields $\Tau^W_m$.
While fitting and marginalizing over $\Tau^W_0$ and $\delta^W$ is an
efficient way of dealing with the super-sample covariance \cite{Li:2014sga,Li:2014jra},
using the power spectrum components of $m>0$ one can directly probe the
super-sample tidal fields, that are usually not directly observable unless
a larger volume containing the current survey is observed.

Our numerical work also shows that $\delta^W$ and linear bias are highly
degenerate, as expected. For measuring tidal fields, we find a shallow
optimum survey size that is given by two competing effects: increasing
volume increases the precision with which we can measure the power spectrum
but at the same time decreases the expected signal of the super-sample
modes. The optimum is at around $V\sim 3\times10^{7}~\hMpcc$ and depends
only weakly on the number density involved. However, in general we find
that the constraint on the tidal fields depends strongly on the galaxy
number density, and for a realistic survey the signal-to-noise ratio is
generally below unity, indicating that it is challenging to probe the
super-sample tidal fields by measuring the anisotropic galaxy power
spectrum. On the other hand, for a high number density galaxy survey
($P_{\rm shot}=1000~\hMpcc$) it is possible to put $\sim1-\sigma$ constraints
on the super-survey tidal fields of $\Tau^W_1$ and $\Tau^W_2$ at their
$\Lambda$CDM expected values. Finally, it is logically possible that
when measured, the measured tidal field would turn up to be considerably
larger than expected, perhaps due to new physics at the horizon scale.
Our result indicates that if the actual tidal fields are no more than
an order of magnitude larger than expected value, they are likely to
be measurable with high significance.

When this technique is applied to a real survey, the curvature of the
sky and shape of the actual window survey would need to be taken into
account carefully. For a fixed large-scale tide, the tidal tensor will
be rotated with respect to the line of sight across a survey covering
large fraction of the sky. A correct methodology for dealing with this
exceeds the scope of this paper.

Another way to probe the super-volume tidal fields is to divide the
entire survey into smaller subvolumes and measure the fully anisotropic
power spectrum in each subvolume, as the position-dependent power spectrum
approach \cite{Chiang:2014oga,Chiang:2015eza}. In this way, one measures
the tidal fields with scale larger than the subvolume size but smaller
than the entire survey, hence the signal-to-noise is expected to be
much higher compared to the super-survey modes.  Since the effect of
the long mode on the large-scale overdensity and tidal fields is
equivalent to the squeezed bispectrum, the same information can also
be extracted from full bispectrum measurements. However, there are
now numerically highly efficient methods for bispectrum measurements
\cite{Schmittfull:2014tca,Scoccimarro:2015bla,Sugiyama:2018yzo} that
make the measurements based on subvolume power spectrum variations
likely obsolete.

\acknowledgments Authors thanks Kazuyuki Akitsu, Donghui Jeong,
Eiichiro Komatsu, Naonori Sugiyama, and Masahiro Takada for helpful
discussion and comments on the draft. We also thank Fabian Schmidt
for pointing out the existence of Poisson modulation term in the
galaxy bispectrum and other useful comments.
CC is supported by grant NSF PHY-1620628.
AS acknowledges hospitality of the Cosmoparticle Hub at University
College London where parts of this work have been performed.

\appendix
\section{Angular decomposition of redshift-space galaxy power spectrum
in the presence of long-wavelength overdensity and tide}
\label{app:Amn}
From \refeqs{delta_resp}{tauij_resp}, it is straightforward to find
\ba
 \:&A_{0,0}=\(b_1^2\)P_l(k) \,, \quad
 A_{0,1}=\(2b_1f\)P_l(k) \,, \quad
 A_{0,2}=\(f^2\)P_l(k) \,, \vs
 \:&A_{1,0}=\(\frac{47}{21}b_1^2+2b_1b_2+\frac{1}{3}b_1^2f\)P_l(k)+\(-\frac{1}{3}b_1^2\)P'_l(k) \,, \vs
 \:&A_{1,1}=\(\frac{26}{7}b_1f+2b_1^2f+2b_2f\)P_l(k)+\(-\frac{2}{3}b_1f-\frac{1}{3}b_1^2f\)P'_l(k) \,, \vs
 \:&A_{1,2}=\(\frac{31}{21}f^2+\frac{10}{3}b_1f^2-\frac{1}{3}f^3\)P_l(k)+\(-\frac{1}{3}f^2-\frac{2}{3}b_1f^2\)P'_l(k) \,, \vs
 \:&A_{1,3}=\(\frac{4}{3}f^3\)P_l(k)+\(-\frac{1}{3}f^3\)P'_l(k) \,, \quad
 A_{2,0}=\(\frac{8}{7}b_1^2+2b_1b_{s^2}\)P_l(k)+\(-b_1^2\)P'_l(k) \,, \vs
 \:&A_{2,1}=\(\frac{24}{7}b_1f+2b_{s^2}f\)P_l(k)+\(-2b_1f\)P'_l(k) \,, \quad
 A_{2,2}=\(\frac{16}{7}f^2\)P_l(k)+\(-f^2\)P'_l(k) \,, \vs
 \:&A_{3,0}=\(-b_1^2f\)P'_l(k) \,, \quad
 A_{3,1}=\(4b_1f^2\)P_l(k)+\(-2b_1f^2\)P'_l(k) \,, \vs
 \:&A_{3,2}=\(4f^3\)P_l(k)+\(-f^3\)P'_l(k) \,, \quad
 A_{4,0}=\(b_1^2f\)P_l(k) \,, \quad
 A_{4,1}=\(-f^3\)P_l(k) \,.
\ea

\bibliography{main}
\end{document}